\documentclass[preprint,usenatbib]{mn2e}
\usepackage{longtable}
\input psfig.sty

\title[Evolutionary Population Synthesis]
{Integrated Spectral Energy Distributions and Absorption Feature Indices of Single
Stellar Populations}

\author[F. Zhang et al.]
{Fenghui~Zhang,$^{1}$\thanks{E-mail: gssephd@public.km.yn.cn} Zhanwen~Han,$^{1}$ Lifang~Li,$^{1}$ Jarrod~R.~Hurley$^{2}$\\
 $^{1}$National Astronomical Observatories/Yunnan Observatory, the Chinese Academy of Sciences, P.O. box 110, Kunming, Yunnan Province, 650011, P.R.China \\
 $^{2}$Department of Astrophysics, American Museum of Natural History, Central Park West at 79th Street, New York, NY 10024, USA
}

\begin{document}

\date{\today}

\pagerange{\pageref{firstpage}--\pageref{lastpage}} \pubyear{2003}

\maketitle

\label{firstpage}

\begin{abstract}
Using evolutionary population synthesis (EPS) we present
integrated spectral energy distributions (ISEDs) and
absorption-line indices defined by the Lick Observatory image
dissector scanner (referred to as Lick/IDS) system, for an
extensive set of instantaneous burst single stellar populations
(SSPs). The ages of the SSPs are in the range $1 \leq \tau /{\rm
Gyr} \leq 19$ and the metallicities $-2.3 \leq {\rm [Fe/H]} \leq
+0.2$. Our models use the rapid single stellar evolution (SSE)
algorithm of \citet*{hur2000} for the stellar evolutionary tracks,
the empirical and semi-empirical calibrated BaSeL-2.0 model of
\citet*{lej97b,lej98} for the library of stellar spectra and
empirical fitting functions of \citet{wor94b} for the Lick/IDS
spectral absorption feature indices.

Applying our synthetic Lick/IDS absorption-line indices to the
merit function we obtain the age and the metallicity of the
central region of M32, it can be well interpreted with an
instantaneous SSP with an age of $\sim$ 6.5\,Gyr and a metallicity
similar to solar. Applying the derived age and the metallicity
from the merit function to a number of index-index diagrams, we
find that the plots of $H_\beta - Fe5015$ and $H_\beta - Fe5782$
are the best index-index diagrams from which we can directly
obtain reasonable age and metallicity.
\end{abstract}

\begin{keywords}
stars: evolution -- galaxies: individual:M32 -- galaxies: stellar
content.
\end{keywords}

\section{Introduction}
Age effects often mask metallicity effects in the studies of
stellar populations \citep{oco76,oco86,oco94,wor92}, and
separating them is a cumbersome affair \citep*{ren86,buz92,buz93}.
The age-metallicity degeneration originates from the fact that
increasing either the metallicity or the age makes the integrated
spectral energy distribution (ISED) of a single stellar population
(SSP) redder \citep*{bre96}.

Previous studies showed that it is difficult to break this
degeneration only by broad-band colours \citep{ari96,wor94a}. In
order to solve this question, some studies used spectral
information, instead of colours only, in the evolutionary
population synthesis (EPS) models
\citep*{bre94,bru93,bru96,bru2003,buz89,kod97,tan96,vaz99a,wor94a}.
With the development of these models including spectral
information, the spectral resolution has been improved from about
20\,\AA \ to 2\,\AA \ \citep{vaz99a}.

Furthermore, these spectral information has been translated to the
line strengths either by empirical fitting polynomials
\citep{gor93,wor94b,wor97} or approaches other than the Lick
fitting functions \citep{pel89,buz95}. The inclusion of absorption
feature indices in EPS models could add the power of diagnostics
to the study of stellar populations \citep{wor94a}, therefore some
studies have included spectral absorption feature indices into EPS
models attempting to understand the stellar populations. Some
earlier EPS studies contained spectral indices include no
metallicity dependence \citep{bru83,tin72a,tin72b,tin76} or are
otherwise limited in scope \citep*{aar78,fro80,mou78,tri92}.
Recently, the spectral absorption indices have been combined
systematically in EPS studies, in those models some model builders
use the absorption feature indices at intermediate resolution
(9\,\AA) of Lick system
\citep*{bru96,bru2003,jon95,kur99,pel89,vaz96,vaz99a,wor94a},
which has been accepted by many investigators and widely used in
their studies; other model builders use the indices at
considerably higher resolution (FWHM $\sim$ 2\,\AA) in Rose system
\citep{buz92,buz93,gor93,ros94,vaz99a,vaz2001}.

SSPs are assemblies of chemically homogeneous and coeval single
stars, the star formation history of any stellar system can be
described by a superposition of SSP models with different ages and
metallicities. Studying SSPs can help us to understand the
evolution of clusters and galaxies, the distribution of
metallicities, and to quantify the star formation history of the
galaxies. In this paper we also investigate SSPs with a systematic
and self-consistent set of stellar evolution models by
\citet{pol98}, spectral library -- the BaSeL-2.0 model
\citep{lej97a,lej97b,lej98} and the spectral absorption feature
indices of Lick system \citep{gor93,wor94b}.

The outline of the paper is as follows. In Section 2, we describe
our SSPs models and algorithm. In Section 3, we give the results
and discussion. In Section 4, we apply the observed spectral
indices to the merit function to determine the age and the
metallicity for SSP-like assembly. Finally we present the summary
and conclusions in Section 5.

\section{Model description}
\subsection{Input physics}
In this study we use the stellar evolutionary models of
\citet{pol98} obtained with the Eggleton stellar evolutionary code
\citep*{egg71,egg72,egg73,han94,pol95} and the empirical and
semi-empirical calibrated BaSeL-2.0 stellar spectral library of
\citet*{lej97b,lej98}.

Instead of using a tabular form in the previous EPS studies
\citep{sch92,cha93,mow98} we choose the evolutionary models of
\citet{pol98} in the convenient form of the rapid single star
evolution (SSE) package presented by \citet*{hur2000}. Except for
a set of analytic evolution functions fitted to the \citet{pol98}
model tracks, the SSE package extends the tracks by including a
description of the remnant phases of stellar evolution, such as
the white dwarf cooling track, and supplements the models of
\citet{pol98} by including a prescription for mass loss, which has
been neglected by the models of \citet{pol98}. The detailed
descriptions about the stellar evolutionary models of
\citet{pol98}, the SSE package of \citet{hur2000} and the
BaSeL-2.0 stellar spectra library of \citet{lej97b,lej98} have
been presented in \citet[][paper I]{zha2002}, so we do not discuss
them here. We refer the interested reader to parts 2 and 3 of
paper I for them. In this paper the factor $\eta$ (Reimers'
mass-loss coefficient, \citealt{rei75}) is taken to be 1/4
\citep{ren81,ibe83,car96}.

Except for the ISEDs, we will also calculate a series of Lick/IDS
(image dissector scanner) spectral absorption feature indices. The
Lick/IDS indices are the absorption strengths in 'feature'
bandpass, and the 'feature' regions are listed in Table 1 of
\citet{wor94b}. In Lick system 21 absorption features are
included, six different molecular bands (CN4150, G band, MgH, MgH
+ Mgb, and two TiO bands) and 14 different blends of atomic
absorption lines. The indices of molecular bands are expressed in
magnitudes, the atomic features indices are expressed in angstroms
of equivalent width (EW).

The Lick/IDS absorption-line indices are given by a series of
empirical fitting functions of \citet{wor94b} as a function of
$T_{\rm eff}$, ${\rm log} g$, and metallicity [Fe/H]. The
effective temperature spans a range 2100 $ \leq T_{\rm eff}/{\rm
K} \leq $ 11000 and the metallicity is in the range $-1.0 \leq
{\rm [Fe/H]} \leq +0.5$. The indices in the Lick system were
extracted from the spectra of 460 stars obtained between 1972 and
1984 using the red-sensitive IDS and Cassegrain spectrograph on
the 3m Shane telescope at Lick Observatory. The spectra cover the
range $4000-6400$\,\AA, with a resolution of $\sim$ 8\,\AA \
\citep{wor94b}.

\subsection{Model input}
For the EPS of an instantaneous burst SSP the main input model
parameters are: (i) the initial mass function (IMF), which gives
the relative number of stars in different evolutionary stages;
(ii) the lower and upper mass cut-offs $M_{\rm l}$ and $M_{\rm
u}$; (iii) the relative age, $\tau$, of the SSP; and (iv) the
metallicity $Z$ of the stars.

We use the IMF of \citet*[][hereafter KTG]{kro93}. The KTG IMF
takes the form
\begin{equation}
\phi(M)= {\rm A}\cdot \Biggl\{ \matrix{ 0.035M^{-1.3} & \ {\rm
if}\ \  0.08 \leq M < 0.5, \cr 0.019M^{-2.2} & {\rm if}\ \  0.5
\leq M < 1.0, \cr 0.019M^{-2.7} & {\rm if}\ \  1.0 \leq M <
\infty, \cr }
\label{IMF-KTG}
\end{equation}
where A is a normalization constant and $M$ is the stellar mass in
solar units. The relative number of stars in the mass range $M
\rightarrow M+{\rm d}M $ is $\phi(M) {\rm d}M$ where $\phi(M)$ is
normalized by
\begin{equation}
\int_{M_{\rm l}}^{M_{\rm u}}\phi(M)M{\rm d}M = 1 \, .
\label{IMF-nor-Sal}
\end{equation}
Taking the lower and upper mass limits, $M_{\rm l}$ and $M_{\rm
u}$, of the stellar mass range as ${\rm{0.1\,M_\odot}}$ and
${\rm120\,{M_\odot}}$ respectively, gives the normalization
constants $A \simeq 16.4$ for the KTG IMF.

\subsection{Algorithm}
Once the input physics database is given, we obtain evolutionary
parameters such as luminosity $L$, effective temperature $T_{\rm
eff}$, radius $R$ and mass $M$ for each star in a SSP using the
SSE algorithm, transform these evolutionary parameters to stellar
flux with the BaSeL-2.0 stellar spectral model, and obtain
absorption feature indices of the Lick/IDS system using the
fitting functions of \citet{wor94b}. By the following equations
(\ref{sp-lamda} - \ref{inte-mag1}) we can obtain the integrated
monochromatic flux and absorption feature indices for an
instantaneous SSP of a particular age and metallicity.

In the following equations, a parameter identified by a capital
letter on the left-hand side represents the integrated SSP, while
the corresponding parameter in minuscule on the right-hand side is
for stars. The integrated monochromatic flux of a SSP is defined
as
\begin{equation}
F_{\lambda,\tau,Z} = \int_{M_{\rm l}}^{M_{\rm u}} \phi(M) \cdot
f_{\lambda} \cdot {\rm d}M , \label{sp-lamda}
\end{equation}
where $f_{\lambda}$ is the SED of stars with mass $M$ and
metallicity $Z$ at a relative age of $\tau$.

The integrated absorption feature index of the Lick/IDS system is
a flux-weighed one. For the $i-$th atomic absorption line, it is
expressed in equivalent width ($W$, in\,\AA),
\begin{eqnarray}
W_{i,\tau,Z} = {{\int_{M_{\rm l}}^{M_{\rm u}} w_{\rm i} \cdot
f_{i,{\rm C}\lambda} \cdot \phi(M) \cdot {\rm d}M} \over
{{\int_{M_{\rm l}}^{M_{\rm u}} \ f_{i,{\rm C}\lambda} \cdot
\phi(M) \cdot {\rm d}M}}},
\label{inte-EW}
\end{eqnarray}
where $w_{\rm i}$ is the equivalent width of the $i-$th index of
stars with mass $M$ and metallicity $Z$ at a relative age of
$\tau$, and $f_{i,{\rm C}\lambda}$ is the continuum flux at the
midpoint of the $i-$th 'feature' passband; and for the $i-$th
molecular line, the feature index is expressed in magnitude,
\begin{equation}
C_{i,\tau,Z} = -2.5 \ {{\int_{M_{\rm l}}^{M_{\rm u}} 10^{-0.4
c_{\rm i}} \cdot f_{i,{\rm C}\lambda} \cdot \phi(M) \cdot {\rm
d}M} \over {{\int_{M_{\rm l}}^{M_{\rm u}} f_{i,{\rm C}\lambda}
\cdot \phi(M) \cdot {\rm d}M}}},
\label{inte-mag1}
\end{equation}
where $c_{\rm i}$ is the magnitude of the $i-$th index of stars
with mass $M$ and metallicity $Z$ at age $\tau$.

For stars cooler than 3570K, the fitting functions of
\citet{wor94b} use two completely different sets of coefficients
for giants and dwarfs, so we must give a criteria to distinguish
them. We adopt the description of \citet{vaz96} for giants and
dwarfs, i.e.,
\begin{equation}
\Bigl\{
\matrix{
{\rm dwarfs},       & \ {\rm if}\ \ {\rm log} g \geq 4.0, \cr
{\rm giants},       & \ {\rm if}\ \ {\rm log} g \leq 3.5. \cr
}
\label{gd-def}
\end{equation}
while for stars with gravity in the range $3.5 <{\rm log}g<4.0$,
$V-K$ is used to discriminate two classes of stars, i.e.,
\begin{equation}
\Bigl\{
\matrix{
{\rm dwarfs},       & \ {\rm if}\ \ (V-K) \leq {\rm log} g -6, \cr
{\rm giants},       & \ {\rm if}\ \ (V-K) > {\rm log} g -6. \cr
}
\label{IMF-KTG}
\end{equation}

\section{Results and Discussion}
In this part we present the ISEDs at intermediate resolution
(10\,\AA \ in the ultraviolet and 20\,\AA \ in the visible) and
the Lick/IDS absorption-feature indices for SSPs over a large
range of age and metallicity: $1 \leq \tau \leq 19 \,$Gyr and
$-2.3 \leq {\rm [Fe/H]} \leq +0.2$. As a check on our models we
compare our synthetic ISEDs with those of \citet[][hereafter
GW]{wor94a} and that of NGC 6838, and compare our synthetic
Lick/IDS absorption-feature indices with the values of
\citet{wor94a}, \citet[][hereafter VCPB]{vaz96}and
\citet[][hereafter KAF]{kur99}, and those of Galactic and M31
globular clusters.

\subsection{Integrated Spectral Energy Distribution}

\begin{figure}
\psfig{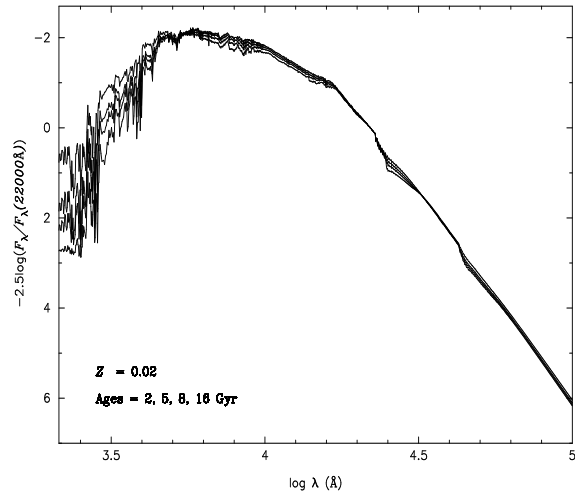}
\caption{The integrated spectral energy distributions as a
function of age for solar-metallicity SSPs. From top to bottom the
age $\tau$= 2, 5, 8 and 16\,Gyr respectively. The flux is
expressed in magnitude and is normalized to zero at 2.2\,$\mu$m.}
\label{ISED-t}
\end{figure}

\begin{figure}
\psfig{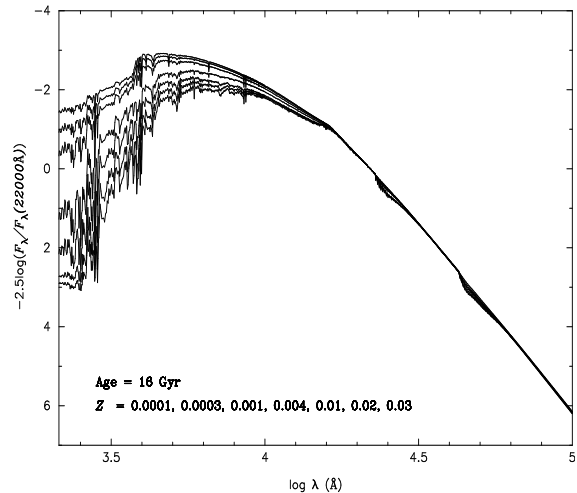}
\caption{The integrated spectral energy distributions as a
function of metallicity for SSPs at age of $\tau$ = 16\,Gyr. From
top to bottom the metallicity $Z$ is 0.0001, 0.003, 0.001, 0.004,
0.01, 0.02 and 0.03 respectively. The flux is also expressed in
magnitude and is normalized to zero at 2.2\,$\mu$m.}
\label{ISED-z}
\end{figure}

\begin{figure}
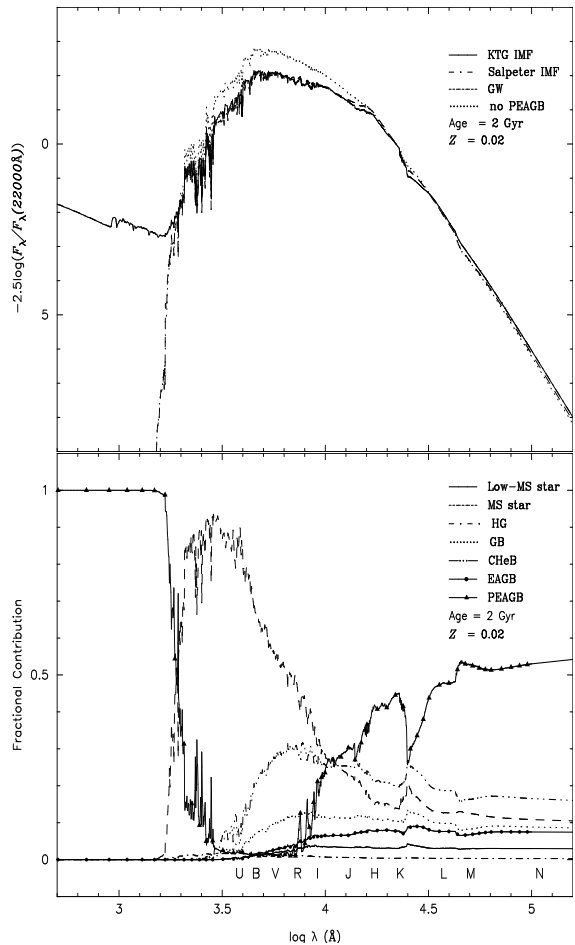

\psfig{file=md897fig31.ps,height=5.9cm,width=7.5cm,bbllx=532pt,bblly=37pt,bburx=82pt,bbury=701pt,clip=,angle=270}
\psfig{file=md897fig32.ps,height=6.5cm,width=7.5cm,bbllx=579pt,bblly=37pt,bburx=82pt,bbury=701pt,clip=,angle=270}
\caption{{\bf In the top panel} comparison of ISEDs for SSPs with
age $\tau=2$\,Gyr and solar metallicity is shown for this work
(solid line) and Worthey (dashed line). In order to analyze the
discrepancy of ISEDs we also give the ISEDs for our SSPs drawn
from the Salpeter IMF with slope $\alpha=2.35$ (dot-dash line, in
the far-ultraviolet region overlapping the ISED with the KTG IMF)
and that for our SSPs drawn from the KTG IMF but without PEAGB
(see the definition in the text) stars included (dotted line, in
the far-ultraviolet region overlapping GW's ISED). The flux is
also expressed in magnitude and is normalized to zero at
2.2\,$\mu$m. {\bf In the bottom panel} the fractional
contributions of different evolutionary stage to the total flux is
shown for corresponding SSPs. The abbreviations are explained in
the text.
}
\label{ISED-com02}
\end{figure}

\begin{figure}
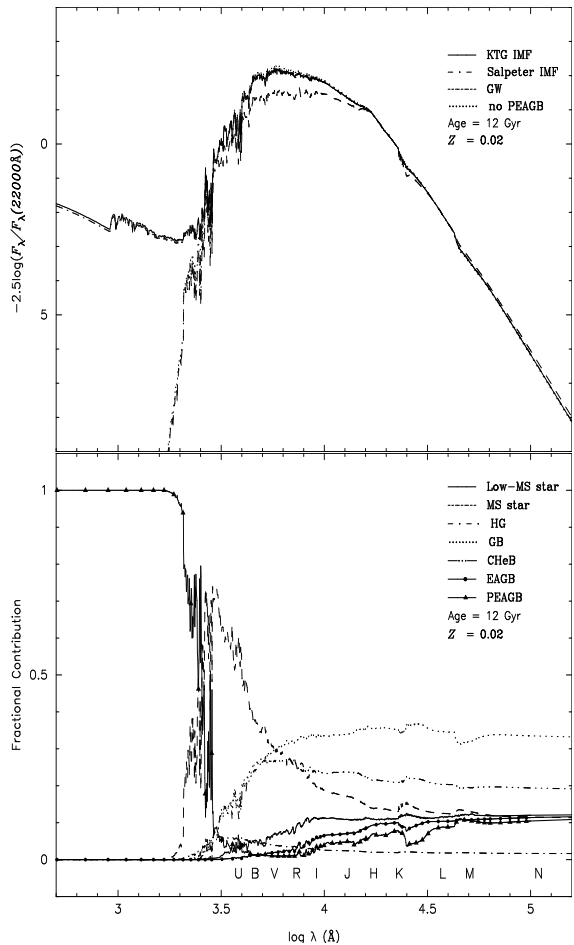

\psfig{file=md897fig41.ps,height=5.9cm,width=7.5cm,bbllx=532pt,bblly=37pt,bburx=82pt,bbury=701pt,clip=,angle=270}
\psfig{file=md897fig42.ps,height=6.5cm,width=7.5cm,bbllx=579pt,bblly=37pt,bburx=82pt,bbury=701pt,clip=,angle=270}
\caption{Similar to Fig. ~\ref{ISED-com02}, but for an age of
12\,Gyr.} \label{ISED-com12}
\end{figure}

\begin{figure}
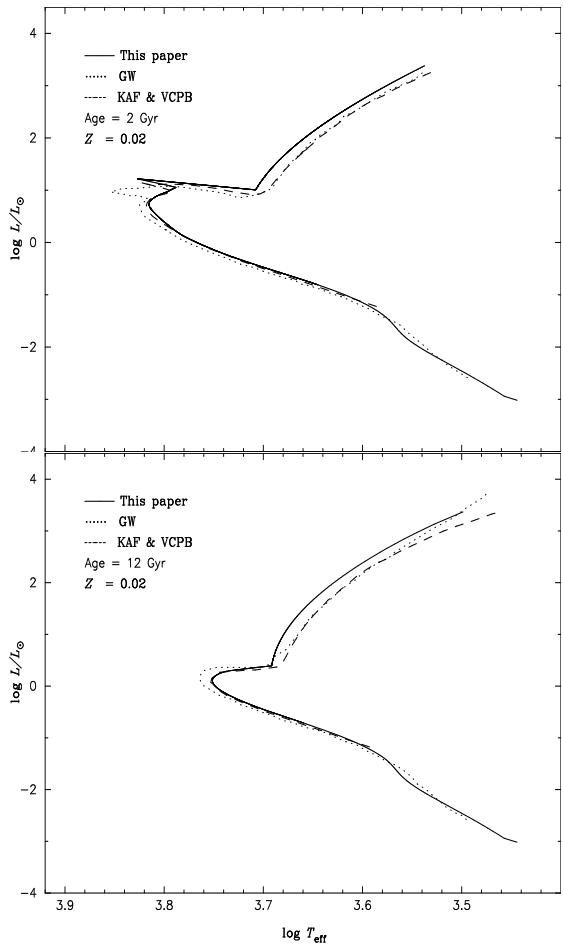

\psfig{file=md897fig51.ps,height=5.9cm,width=7.5cm,bbllx=532pt,bblly=37pt,bburx=82pt,bbury=701pt,clip=,angle=270}
\psfig{file=md897fig52.ps,height=6.5cm,width=7.5cm,bbllx=583pt,bblly=37pt,bburx=82pt,bbury=701pt,clip=,angle=270}
\caption{Theoretical isochrones of 2\,Gyr and 12\,Gyr old,
instantaneous burst stellar populations with solar metallicity.
Results from this work, GW, KAF and VCPB are shown.}
\label{synch-com}
\end{figure}

In Figs. \ref{ISED-t} and \ref{ISED-z} we give the variation of
the intermediate resolution ISED with age and metallicity over a
wide wavelength range, $3.3 \leq {\rm log}(\lambda/{\rm \AA}) \leq
5.0$. The flux is expressed in magnitude and is normalized to zero
at 2.2\,$\mu$m. Figs. \ref{ISED-t} and \ref{ISED-z} show that the
effects of age and metallicity on the ISEDs are similar, i.e., the
ISEDs tend to be redder with increasing age or metallicity in the
wavelength region of $3.3 \leq {\rm log} (\lambda/{\rm \AA}) \leq
4.3$.

\subsubsection{Comparison with the results of previous studies}
To test our models we compare our synthetic ISEDs with those of GW
for the young ($\tau=2\,$Gyr) and the old ($\tau=12\,$Gyr) SSPs
with solar-metallicity in the top panels of Figs. \ref{ISED-com02}
and \ref{ISED-com12}, respectively. The ISEDs of GW are based on
his work in 1994 and from his private homepage \citep{wor2002},
which allows one to obtain ISEDs and the Lick/IDS absorption
feature indices for arbitrary mixtures of single-burst SSPs with
ages in the range $1 \leq \tau \leq 18\,$Gyr and metallicities in
the range $-2 \leq {\rm [Fe/H]} \leq +0.5$ (but not for
metallicity ${\rm [Fe/H]} < -0.225$ with age $\tau < 8\,$Gyr).

Figs. \ref{ISED-com02} and \ref{ISED-com12} show significant
disagreement of the ISED in two wavelength regions, the larger one
is in the far-ultraviolet region (${\rm log} (\lambda/{\rm \AA}) <
3.3$), the minor is in the visible and infra-red regions (i.e.,
$3.5 < {\rm log} (\lambda/{\rm \AA}) < 4.2$). In the first
wavelength region our ISED is completely different from that of
GW, in the second region this work exhibits bluer continuum than
that of GW. Comparing the discrepancy of ISED in the visual and
infra-red regions for the young and the old SSPs it seems that
this discrepancy in ISEDs increases with age.

\vskip 0.7cm \leftline{( 1 ) \it{\ Model Comparison}} \vskip 0.2cm

In order to investigate what actually causes the discrepancy
between our ISEDs and those of GW, we first discuss the
differences between these two EPS models. This work differs from
that of GW by the adoption of
\begin{itemize}
\item different stellar evolutionary models. We use
the stellar evolutionary models of \citet{pol98}, whereas GW use
the stellar evolutionary isochrones by VandenBerg and
collaborators \citep{van85a,van92,van85b,van87} and the revised
Yale isochrones \citep{gre87}. And in the GW model the post
asymptotic giant branch (AGB) stars are not included.
\item different stellar spectral libraries. We use the BaSeL-2.0
library, which is based on several original grids of model
atmosphere spectra: Kurucz (1995, private communication) for O to
late -K stars; \citet{ful94} and \citet{bes89,bes91} for M giants
in the temperature range $3500 - 2500\,$K; and \citet{all95} for M
dwarfs, whereas GW used the theoretical spectra of \citet{kur92}
for stars hotter than 3750K and composite spectra for cooler stars
by patching together model atmospheres (\citealt{bes89,bes91}, and
references therein) and optical observational spectra of M giants
\citet{gun83}.
\item different IMF shapes. We use the KTG IMF
\citep{kro93}, whereas GW adopt the Salpeter IMF with a slope of
$\alpha = 2.35$ \citep{sal55}, i.e.,
\begin{equation}
\phi(M) = {\rm C} \cdot M^{-2.35} \ \ \
\label{IMF-Sal}
\end{equation}
where C is a normalization constant.
\end{itemize}

The difference between our ISEDs and those of GW can possibly be
attributed to either the choice of stellar evolutionary models,
the spectral library, the IMF and/or the inclusion of post AGB in
EPS model. In the top panels of Figs. \ref{ISED-com02} and
\ref{ISED-com12} we supplement the ISED drawn from the Salpeter
IMF with slope $\alpha = 2.35$ and the ISED from the KTG IMF but
no thermally pulsing giant branch/proto planetary nebula/planetary
nebula (TPAGB/PPN/PN, these phases are beyond "early asymptotic
giant branch [EAGB]", are collectively termed "post EAGB [PEAGB]")
stars included for solar metallicity SSPs at ages of 2\,Gyr and
12\,Gyr. In the bottom panels we give the fractional contributions
of different evolutionary stage to the total flux for 2\,Gyr and
12\,Gyr SSPs of solar metallicity. In it various abbreviations are
used to assign the evolution phases. They are as follows: "MS"
stands for main-sequence stars, "HG" stands for Hertzsprung gap;
"GB" stands for the first giant branch; "CHeB" stands for core
helium burning; and the MS is divided into two phases to
distinguish deeply or fully convective low-mass stars ($M <
{\rm{0.7\,M_\odot}}$) and stars of higher mass with little or no
convective envelope ($M \geq {\rm{0.7\,M_\odot}}$).


\vskip 1.0cm \leftline{( 2 ) \it{\ Analysis of the Difference of
ISEDs in the}} \leftline{\it\,\ \ \ \ \ \ \ far-ultraviolet
Region} \vskip 0.2cm

First, the adoption of different spectral libraries can not cause
significant discrepancy of ISEDs in the region ${\rm log}
(\lambda/{\rm \AA}) < 3.3$. The reason is that the total light in
this wavelength region is dominated by PEAGB stars (see the bottom
panels of Figs. \ref{ISED-com02} and \ref{ISED-com12}), in an
exact word, by PN stars with temperature ${\rm log} (T_{\rm
eff}/{\rm K}) > 5.0$. For these hotter PN stars the GW model and
our model actually use the same spectral library.

Furthermore, the top panels of Figs. \ref{ISED-com02} and
\ref{ISED-com12} show that (i) the ISED of GW in the
far-ultraviolet region agree with our result with the KTG IMF but
no PEAGB stars included ; (ii) two ISEDs overlap for our SSPs with
the different IMF shapes, i.e., the KTG IMF and the Salpeter IMF
with a slope of $\alpha = 2.35$ in the region ${\rm log}
(\lambda/{\rm \AA}) < 3.3$.

Therefore the significant disagreement between our ISED in the
region ${\rm log} (\lambda/{\rm \AA}) < 3.3$ and that of GW is
dominated by the inclusion of PN stars in PEAGB stage for SSPs,
and the IMF shape is not the main factor.

\vskip 0.7cm
\leftline{( 3 ) \it{\ Analysis of the Difference in
the Visible and}}
\leftline{\it\,\ \ \ \ \ \ \ Infra-red Region}
\vskip 0.2cm

From the bottom panel of Fig. \ref{ISED-com02} we see that the
total flux in the visible and infra-red regions is dominated by MS
stars with mass $M \geq {\rm{0.7\,M_\odot}}$, CHeB stars and
cooler TPAGB/PPN stars in PEAGB stage (only at wavelength ${\rm
log} (\lambda/{\rm \AA})> 3.9$) for young SSPs ($\tau = 2\,$Gyr).
The contribution of MS stars to the light is greater than those of
CHeB and the cooler TPAGB/PPN stars in the region of 3.5 $ < {\rm
log} (\lambda/{\rm \AA}) < $ 4.0, while the cooler TPAGB/PPN stars
make the greatest contribution to the light in the region 4.0 $ <
{\rm log} (\lambda/{\rm \AA}) < $ 4.2.

From the bottom panel of Fig. \ref{ISED-com12} we see that the
light in the visible and infra-red regions is dominated by MS
stars with mass $M \geq {\rm{0.7\,M_\odot}}$, CHeB stars and GB
stars for old SSPs ($\tau = 12\,$Gyr). The contribution of MS
stars and GB stars is greater than that of the other two
evolutionary stages at shorter and longer wavelengths,
respectively, and the contribution of MS stars to the light is
equal to GB at wavelength ${\rm log} (\lambda/{\rm \AA}) \sim $
3.8.

The top panels of Figs. \ref{ISED-com02} and \ref{ISED-com12} show
that (i) the inclusion of TPAGB/PPN stars in PEAGB stage in EPS
models can produce a significant deviation of ISEDs in the visible
and infra-red regions for the young SSPs ($\tau = 2\,$Gyr), but
almost no effect for the old SSPs ($\tau = 12\,$Gyr). The reason
is that the cooler TPAGB/PPN stars in PEAGB stage contribute a lot
to the light for young SSPs at the red end of this region, but
almost no light contribution for old SSPs (see the bottom panels
of Figs. \ref{ISED-com02} and \ref{ISED-com12}). The relatively
greater contribution of the cooler TPAGB/PPN stars, for young
SSPs, is mainly caused by the fact that the relative number of the
TPAGB/PPN stars is greater than that for old SSPs. (ii) For $\tau
= 2\,$Gyr SSPs, the ISEDs for SSPs without PEAGB stars included
are as much as $\sim$ 1\,mag greater than those for SSPs including
PEAGB stars. This reason of the blueness of ISEDs is that the SSPs
would look like younger and the ISEDs trend to be bluer if
omitting those cooler TPAGB/PPN stars in PEAGB stage. (iii) Also
for $\tau = 2\,$Gyr SSPs, the discrepancy in ISEDs arising from
the inclusion of TPAGB/PPN stars is larger than that arising from
the different models (the GW model and our model). (iv) the
adoption of different IMF shapes almost do not affect the ISEDs in
the visible and infra-red regions. This can be inferred from the
uniqueness of ISEDs between the values with the KTG IMF and those
with the Salpeter IMF with slope $\alpha = 2.35$.

Besides the inclusion of PEAGB stars, the adoption of different
spectral library and evolutionary models can result in the
deviation of ISEDs in the visible and infra-red regions. In Fig.
\ref{synch-com} we plot theoretical isochrones for
solar-metallicity SSPs at ages of 2\,Gyr and 12\,Gyr, as
calculated by GW and this work. It reveals three significant
deviations between the two isochrones: (i) MS stars with mass $M
\geq {\rm{0.7\,M_\odot}}$ (corresponding roughly to ${\rm log}
T_{\rm eff}$ of 3.6) in our models are lightly cooler than that in
the GW models; (ii) sub-giant stars in our models are
significantly bluer. The cooler MS stars in our models cause
redder visible and infra-red continuum, the hotter GB stars cause
bluer continuum, and the reddening of ISEDs caused by cooler MS
stars is compensated by the hotter GB stars.

In summary the discrepancy between our ISEDs in the visible and
infra-red regions and those of GW is not caused mainly by the
adoption of different IMF shapes, i.e., the KTG IMF and the
Salpeter IMF with a slope of $\alpha=2.35$, but by the adoption of
different stellar evolutionary models and spectral library in EPS
model. The inclusion of the cooler TPABB/PPN stars in PEAGB stage
can produce a significant deviation of ISEDs for young SSPs, but
almost no effect for old SSPs in this region.

\subsubsection{Comparison with the ISED of NGC 6838}
\begin{figure}
\psfig{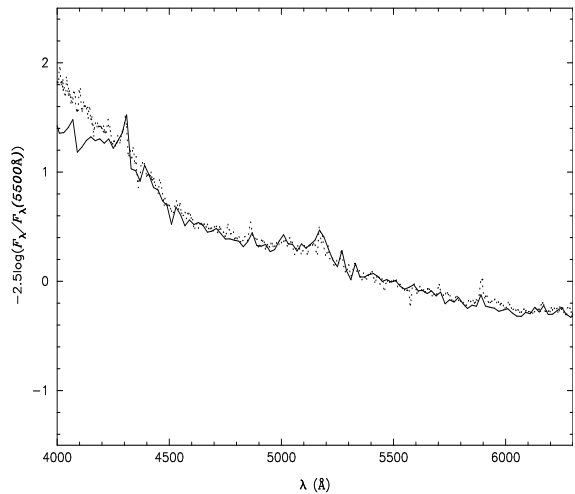}
\caption{The fit to the ISED of NGC 6838 (dotted line) in the
range 4000-6400\,\AA. The model ISED is for a SSP with solar
metallicity and 12.6\,Gyr (solid line). The flux is expressed in
magnitude and is normalized to zero at 5500\,\AA.}
\label{ISED-NGC6838}
\end{figure}

We compare model ISEDs with that of NGC 6838. The spectral data of
NGC 6838 is from \citet{tra98} and covers 4000-6400\,\AA, its
resolution is 1.25\,\AA. The result reveals that NGC 6838 can be
fitted by a SSP with solar metallicity and an age of $\sim$
12.6\,Gyr. In Fig. \ref{ISED-NGC6838} we present the ISED of NGC
6838 and the model ISED for an SSP with solar metallicity and an
age of $\sim$ 12.6\,Gyr, the flux is expressed in magnitude and is
normalized to zero at 5500\,\AA. From Fig. \ref{ISED-NGC6838} we
see that the agreement between model and observed spectrum is
quite well in the range $\lambda < 4300 {\rm \AA}$, but the
agreement is not so good for shorter region.

\subsection{Lick/IDS absorption feature indices}
In Table 1 we present all resulting Lick/IDS spectral absorption
feature indices for 7 metallicities from $Z=0.0001$ to $Z=0.03$.
All indices except for $H_{\beta}$ increase with increasing age
and metallicity, and the variation of these indices with age is
larger at early age.

\subsubsection{Comparison with the results of previous studies}
\begin{figure*}
\psfig{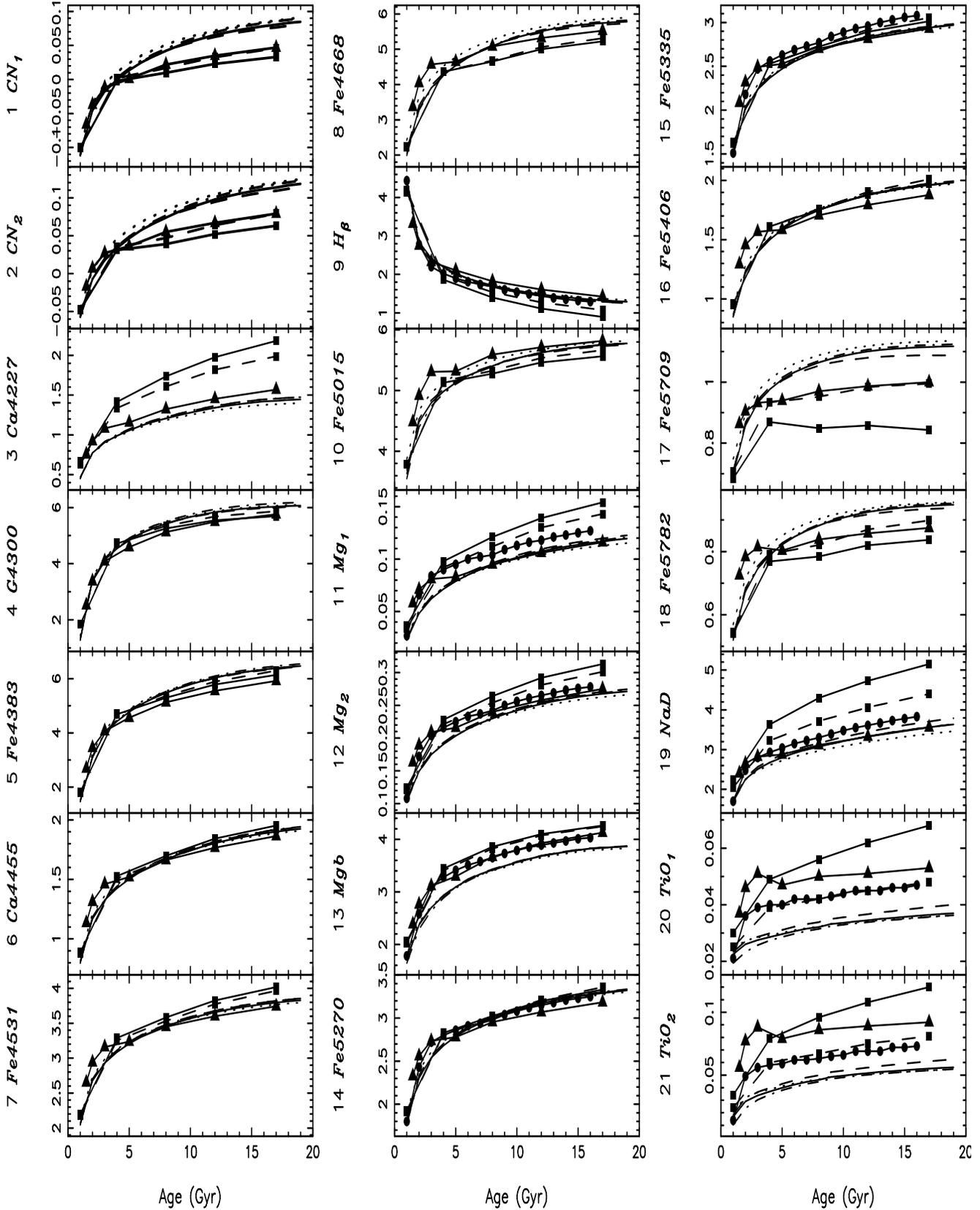}
\caption{Evolution of absorption indices in Lick/IDS system for
solar-metallicity instantaneous burst SSPs, according to several
recent EPS models (Worthey [1994, solid line + solid triangle],
Vazdekis et al. [1996, solid line + solid square for the unimodal
IMF, and dashed-line + solid square for the bimodal IMF], and
Kurth, Alvensleben \& Fricke [1999, solid line + solid circle]).
Indices for our SSPs drawn from the Salpeter IMF with slopes
$\alpha = 2.35$ [dashed line] and 1.35 [dot-dash], and the KTG IMF
but no PEAGB stars included [dotted line] are supplemented.}
\label{asi-com-auth}
\end{figure*}

In Fig. \ref{asi-com-auth} we give the evolution of the Lick/IDS
spectral absorption feature indices for solar-metallicity SSPs
with the KTG IMF, supplemented with the indices for
solar-metallicity SSPs obtained by GW (1994), VCPB (1996) and KAF
(1999). Except for $CN_{\rm 1}$, $CN_{\rm 2}$, $Ca4277$, $NaD$,
$TiO_{\rm 1}$ and $TiO_{\rm 2}$ indices, which show much stronger
discrepancies between the studies, all indices agree with those of
other studies.

\vskip 0.7cm
\leftline{( 1 ) \it{\ Description of the VCPB and the
KAF model}}
\vskip 0.2cm

The VCPB model adopts the theoretical isochrones of \citet{ber94},
empirical spectra whenever possible, the fitting functions for
$CN_{\rm 1}$ and $CN_{\rm 2}$ indices from \citet{vaz96} and the
fitting functions for the other Lick indices from \citet{wor94b},
and two IMF shapes: unimodal and bimodal. The unimodal IMF takes
the from
\begin{equation}
\phi(M) = \beta M^{-\mu}
\label{vaz-imf-1}
\end{equation}
where $\mu$ for the solar neighborhood is equal to 1.35 and
$\beta$ is a constant. The bimodal IMF is described by
\begin{equation}
\phi(M)= \beta \cdot
\Biggl\{
\matrix{
0.4^{-\mu} & \ \ \  M \leq 0.2, \cr
p(M) & \ \ \ \ \ \ \ \ \ \  0.2 < M < 0.6, \cr
M^{-\mu}   & \ \ \  M \geq 0.6, \cr
}
\label{vaz-imf-2}
\end{equation}
where $p(M)$ is a spline. For stars with mass lower than
${\rm{0.6\,M_\odot}}$ the tracks are from \citet{pol98}, as used
by us.

KAF used the evolutionary tracks from the Padova group
\citep{bre93,fag94a,fag94b,fag94c}, for lower masses ($0.08 \leq M
< 0.5\,{\rm{M_\odot}}$) used the tracks of \citet{cha97},
theoretical colour calibrations from \citet{lej97b,lej98}, fitting
functions for stellar atmospheric indices from \citet{wor94b}, and
a Salpeter-like IMF with a slope of 1.35 \citep{sal55}.

\vskip 0.7cm
\leftline{( 2 ) \it{\ Analysis of the Disagreement of
Indices amongst}} \leftline{\ \ \ \ \ \ \ \it{\  those Models}}
\vskip 0.2cm

In order to investigate which factors cause the difference between
our synthetic Lick/IDS absorption-line indices and those of GW,
especially in the $CN_{\rm 1}$, $CN_{\rm 2}$, $Ca4277$, $NaD$,
$TiO_{\rm 1}$ and $TiO_{\rm 2}$ indices, in Fig.
\ref{asi-com-auth} we also give the Lick/IDS absorption feature
indices for SSPs with the Salpeter IMF with a slope of
$\alpha=2.35$ and the indices for SSPs with the KTG IMF but no
PEAGB stars included in EPS model. It shows that the choice of the
IMF and the exclusion of PEAGB stars give rise to only small
changes in $Fe5406$, $NaD$, $TiO_{\rm 1}$ and $TiO_{\rm 2}$
indices, and the deviations introduced by the IMF shape and the
exclusion of PEAGB stars are smaller than the discrepancies
between the two models. So the disagreement in the Lick/IDS
absorption-line indices between our work and the GW model is
mainly caused by the adoption of spectral library and stellar
evolutionary models. The effects of adopting different stellar
evolutionary models on the Lick/IDS indices are as follows: (i)
Cooler MS stars with mass $M \geq {\rm{0.7\,M_\odot}}$ in our
models (see Fig. \ref{synch-com}) make the absorption-line indices
at redder end ($NaD$, $TiO_{\rm 1}$, $TiO_{\rm 2}$) and the blue
index $Ca4277$ higher because these indices decrease sharply with
increasing temperature, but make $CN_1$ and $CN_2$ lower. (ii)
Hotter GB stars in our models (see Fig. \ref{synch-com}) make
$NaD$, $TiO_{\rm 1}$ and $TiO_{\rm 2}$ and $Ca4277$ significantly
lower, but $CN_{\rm 1}$ and $CN_{\rm 2}$ higher. (iii) The effect
of cooler MS stars on the Lick/IDS indices is contrary to that of
hotter GB stars, and the deviations are dominated by MS stars, the
reason is the Lick/IDS indices cover the region 4000 $\sim$
6000\,\AA (i.e., 3.6 $\la {\rm log}(\lambda/{\rm \AA}) \la $ 3.8),
in which the total light is dominated by MS stars with mass $M
\geq {\rm{0.7\,M_\odot}}$.

KAF (1999) use the same spectral library (i.e., the BaSeL-2.0
library) and the same fitting functions for Lick/IDS absorption
indices (i.e. the functions of Worthey et al., 1994) as those in
our model. So the disagreement between our absorption-line indices
and those of KAF is possibly introduced by the adoption of
different evolutionary models and the IMF shape. In Fig.
~\ref{asi-com-auth} we give the indices for SSPs with the
Salpeter-like IMF with a slope of $\alpha=1.35$, as used by KAF.
It shows that the deviation arising from the IMF shape is smaller
than that arising from the different models. So the difference
between this work and the KAF models is mainly caused by the
choice of stellar evolutionary models. In Fig. ~\ref{synch-com}
the theoretical Padova isochrones used by KAF are given for
solar-metallicity SSPs at ages of 2\,Gyr and 12\,Gyr. It shows
that the tracks of MS stars with $M \geq {\rm{0.7\,M_\odot}}$ in
our models agree with those in the KAF models. According the
analysis in the above paragraph, i.e., the deviations are
dominated by MS stars with mass $M \geq {\rm{0.7\,M_\odot}}$, we
can draw the following conclusion: the deviations in the Lick/IDS
indices between our models and KAF models are smaller than those
between our models and GW models, in fact, this conclusion
coincide with the comparison of $TiO_1$ and $TiO_2$ in Fig.
~\ref{asi-com-auth}.

VCPB (1994) use different stellar evolutionary models, spectral
library and two different IMF shapes from those in our models. In
Fig. \ref{asi-com-auth} we give the indices from the VCPB models
for the unimodal and bimodal IMF with a slope of 2.35. It shows
that the significant discrepancies between the unimodal and
bimodal models exist in $Fe5709$, $NaD$, $TiO_{\rm 1}$ and
$TiO_{\rm 1}$ indices. Our results are much closer to those with
the bimodal IMF than those with the unimodal IMF. The differences
between the results in our models and those in the VCPB models are
possibly caused by the adoption of different stellar evolutionary
models, spectral library and the IMF shape. In fact, the VCPB and
KAF models adopt the same isochrones, i.e., Padova isochrones (see
Fig. \ref{synch-com}), so the deviations of Lick/IDS indices
introduced by the different evolutionary models between our and
VCPB models, is same as those between our models and KAF models.

\subsubsection{Comparison with observed clusters}
\begin{figure*}
\psfig{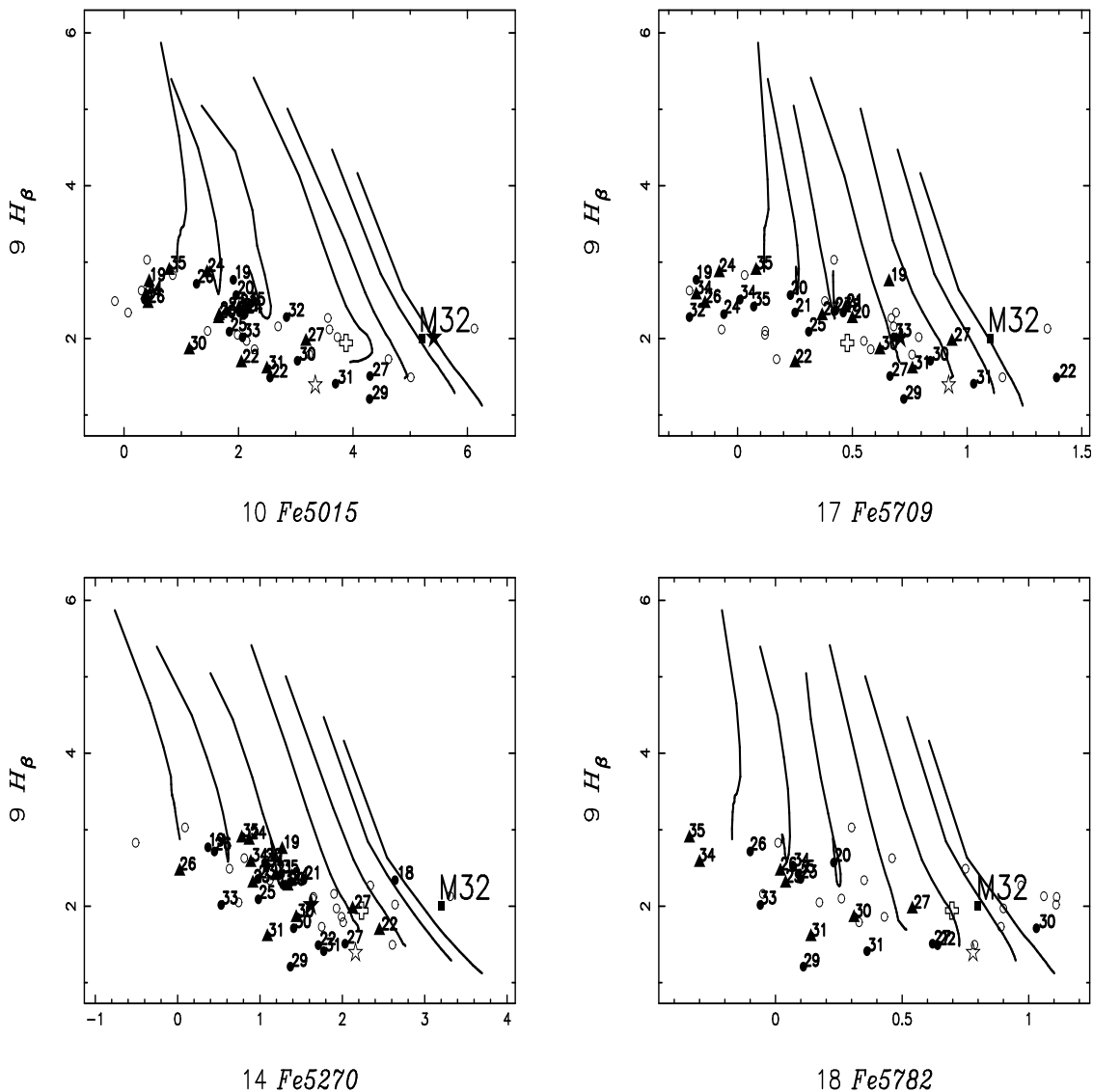}
\caption{Comparison of our integrated indices with observations of
Galactic (solid symbols) and M31 (open circle) globular clusters
and M32(solid square). The number indicate the corresponding
cluster in Tab. \ref{GC-name}. For Galactic cluster, the data for
the center ('C') is indicated by solid circle, those for the
off-center ('O') by solid triangle. For Galactic cluster, NGC
6624, the data of 'O', 'L' and 'S' is indicated by solid star,
open cross and open star, respectively.} \label{asi-com-obs}
\end{figure*}

\setcounter{table}{1}
\begin{table}
\caption{Manual of Observational data}
\begin{tabular}{rrrrrrr}
  \hline
  \hline
  & NAME &  SA  &  C  &  O   &   L  &  S  \\
  \hline
 1    &   M31 4      &  SA  &     &     &    &   \\
 2    &   M31 23     &  SA  &     &     &    &   \\
 3    &   M31 42     &  SA  &     &     &    &   \\
 4    &   M31 76     &  SA  &     &     &    &   \\
 5    &   M31 87     &  SA  &     &     &    &   \\
 6    &   M31 95     &  SA  &     &     &    &   \\
 7    &   M31 99     &  SA  &     &     &    &   \\
 8    &   M31 100    &  SA  &     &     &    &   \\
 9    &   M31 116    &  SA  &     &     &    &   \\
 10   &   M31 196    &  SA  &     &     &    &   \\
 11   &   M31 282    &  SA  &     &     &    &   \\
 12   &   M31 301    &  SA  &     &     &    &   \\
 13   &   M31 MII    &  SA  &     &     &    &   \\
 14   &   M31 MIV    &  SA  &     &     &    &   \\
 15   &   M31 V101   &  SA  &     &     &    &   \\
 16   &   M31 V12    &  SA  &     &     &    &   \\
 17   &   M31 V64    &  SA  &     &     &    &   \\
 18   &   NGC 2158   &      &  C  &     &    &   \\
 19   &   NGC 5024   &      &  C  &  O  &    &   \\
 20   &   NGC 5272   &      &  C  &  O  &    &   \\
 21   &   NGC 5904   &      &  C  &  O  &    &   \\
 22   &   NGC 6171   &      &  C  &  O  &    &   \\
 23   &   NGC 6205   &      &  C  &  O  &    &   \\
 24   &   NGC 6218   &      &  C  &  O  &    &   \\
 25   &   NGC 6229   &      &  C  &     &    &   \\
 26   &   NGC 6341   &      &  C  &  O  &    &   \\
 27   &   NGC 6356   &      &  C  &  O  &    &   \\
 28   &   NGC 6624   &      &     &  O  &  L & S \\
 29   &   NGC 6637   &      &  C  &     &    &   \\
 30   &   NGC 6712   &      &  C  &  O  &    &   \\
 31   &   NGC 6838   &      &  C  &  O  &    &   \\
 32   &   NGC 6981   &      &  C  &     &    &   \\
 33   &   NGC 7006   &      &  C  &     &    &   \\
 34   &   NGC 7078   &      &  C  &  O  &    &   \\
 35   &   NGC 7089   &      &  C  &  O  &    &   \\
 36   &   M32        &      &     &     &    &   \\
  \hline
\end{tabular}
\label{GC-name}
\end{table}

As another test, the model indices are checked against a wide
variety of available observations. We compare our absorption-line
indices with those of Galactic and M31 globular clusters and those
of the circumnuclear region of M32 in many index-index diagrams
(210 diagrams in total, not included in this paper). The line
strengths of Galactic and M31 globular clusters are from
\citet{tra98} and that of M32 from \citet[][references therein
]{del2001}. In the database of \citet{tra98} 36 globular clusters
(18 M31 and 18 Galactic clusters) are included (see Table
\ref{GC-name}). M31 globular clusters are observed with the
standard aperture (1.4\arcsec $\times$ 1.4\arcsec, hereinafter
'SA'). The Galactic globular clusters are observed with a long
slit of standard width (1.4\arcsec $\times$ 16\arcsec) that was
raster-scanned on the sky to create a square aperture of size
66\arcsec $\times$ 66\arcsec, this resulted in two square
apertures, one centered on the cluster (hereinafter 'C') and one
for the "off" beam (hereinafter 'O') located 35\arcsec \ to the
east. For NGC 6624, the 'L' aperture is 45\arcsec \ by 60\arcsec \
and the 'S' aperture is 13\arcsec \ by 13\arcsec, both are
centered on the cluster. Raster scans have the same spectral
resolution as standard-slitwidth scans (1.4\arcsec). In this paper
M31 V204 has been omitted from the database of \citet{tra98}
because of larger deviation from our models.

The results reveal that our models fit the data very well in many
index-index diagrams, while in some index-index diagrams the
models do not match real globular clusters well, e.g. in the plots
of $TiO_{\rm 1}$ vs. other indices. The index $TiO_{\rm 1}$ is
systematically smaller than observations. In Fig.
\ref{asi-com-obs} we only give several representative plots in
which the model predictions fit the data well, i.e., $H_{\beta}$
vs. four iron lines ($Fe5015$, $Fe5270$, $Fe5709$ and $Fe5782$).
From Fig. \ref{asi-com-obs} we see that there are no significant
discrepancies in spectral indices between M32 and Galactic
globular clusters, and the required metallicity is always lower
than solar and the age is older than 10\,Gyr for most of Galactic
and M31 globular clusters. Meanwhile from Fig. \ref{asi-com-obs}
we see that the metallicity of the central region of M32
approaches to solar metallicity.

\section{Determination of age and metallicity}
\subsection{Merit Function}
Our synthetic spectral absorption
indices agree better with the values obtained by other authors and
fit the observational data very well in most of index-index
diagrams, therefore we use these spectral absorption indices to
constrain or determine the age and the metallicity for SSP-like
assemblies. The method is based on minimizing a chi-squared merit
function $F({\rm J})$ between all observed indices and model
predictions. The merit function is a measure for the goodness of
the fit for observational data. For each model $J$ merit function is
defined as:
\begin{equation}
F({\rm J}) = {{\sum_{{\rm i=1}}^{{\rm n}} W_{\rm i} ( {{O_{\rm
i}-M_{\rm i}({\rm J})} \over {E_{\rm i}}} )^2 } \over {{\sum_{{\rm
i=1}}^{{\rm n}} W_{\rm i}}}}
\end{equation}
where $n$ is the number of observed spectral indices, $O_{\rm i}$
is the observed value of the $i-$th index, $M_{\rm i}({\rm J}) $
is the synthetic value of corresponding index for model $J$, and
$E_{\rm i}$ is its observational error, $W_{\rm i}$ represents its
relative weight. In this work we assign the same weight (=1) to
all the spectral indices.

\subsection{Age and metallicity of M32}
\begin{figure}
\psfig{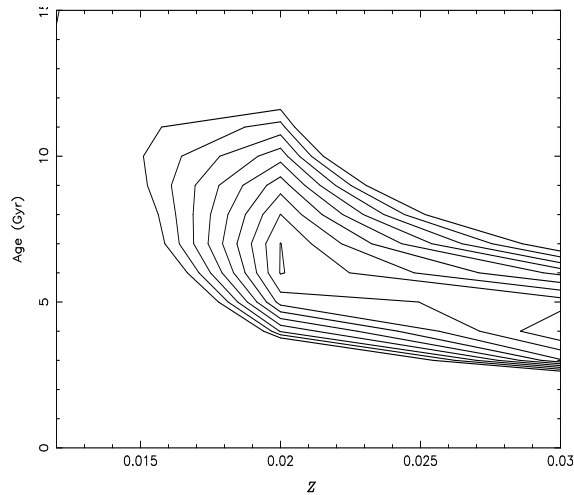}
\caption{Contours of the merit function of M32. From center to
edge the value of $F$ increase from 1.2 to 2.8 at an interval of
0.2.}
\label{contour}
\end{figure}

Using the above merit function $F$ we will study the age and the
metallicity of the central region of M32, which is an elliptical
galaxy. The reason we discuss M32 is as follows. Firstly, M32 is
the brightest elliptical galaxy of the Local Group, and has been
studied extensively as a template for elliptical galaxies further
away. Secondly, the elliptical galaxies can be viewed as SSP-like
assemblies because they appear to contain relatively little dust
extinction and gaseous interstellar medium \citep{vaz96}, and in
them most of the stars were formed during, or very early after the
initial collapse of the galaxy \citep{bru2003}. So we can
investigate it by the SSPs models.

In Fig. \ref{contour} we give its contour map of the merit
function $F$ as a function of metallicity $Z$ and age $\tau$. It
shows that the data of absorption indices can be successfully
fitted with an instantaneous SSP with an age of $\sim$ 6.5\,Gyr
and a solar-like metallicity, with $F$ of 1.00 for the central
region of M32.

The age obtained by us is greater than 4\,Gyr obtained by
\citet{del2001} from their integral field spectroscopy of the 9x12
${\rm arcsec}^2$ circumnuclear region of M32 using 30 spectral
indices and colours and by \citet{vaz99b} from the blue spectrum
using index $H_\gamma$, while younger than 7\,Gyr obtained by
\citet{jon95} based on their ${\rm H}_{\gamma HR}$ index and
8.5\,Gyr from \citet{gri96} based on their HST analysis resolving
M32 into individual stars.

The resulted metallicity is in full agreement with that of
\citet{del2001} and \citet{vaz99b}, both of them predicted
solar-like metallicity, but differ from that of \citet{gri96}, who
predicted a metallicity of ${\rm [Fe/H]} = -0.25$.

Applying the derived age and metallicity of the central region of
M32 from merit function to the index-index diagrams (see Fig.
\ref{contour}), it seems that the plots of $H_\beta - Fe5015$ and
$H_\beta - Fe5782$ are better diagrams to estimate directly the
age and the metallicity from index-index plots. The plots of
$H_\beta - Fe5015$ and $H_\beta - Fe5782$ give the metallicity of
$Z \ga 0.03$ for the center region of M32.

\section{Summary and Conclusion}
In this paper we use the EPS technique to present the ISEDs at
intermediate resolution (10\,\AA \ in the ultraviolet and 20\,\AA
\ in the visible) and the Lick/IDS spectral absorption indices for
an extensive set of instantaneous burst SSPs of various ages and
metallicities ($1 \leq \tau \leq 19\,$Gyr, $-2.3 \leq {\rm [Fe/H]}
\leq 0.2$). In our EPS models we adopted the rapid SSE algorithm,
the empirical and semi-empirical calibrated BaSeL-2.0 spectral
library and the fitting functions for the absorption-line indices.

As checks on our models, we compare our ISEDs with those of
Worthey and the high resolution spectrum of NGC 6838, also compare
the Lick/IDS absorption feature indices with the values of other
EPS models and those of Galactic and M31 globular clusters and the
central region of M32. We conclude that: (a) The disagreement
between our ISEDs with those of Worthey presents in two wavelength
ranges, one in the far-ultraviolet region, the other in the
visible and infra-red regions. The discrepancy of ISEDs in the
far-ultraviolet region is mainly caused by the inclusion of hotter
PN stars in our models, while the discrepancy of ISEDs in the
visible and infra-red regions is caused by the choice of stellar
evolutionary models and spectral library. The inclusion of cooler
TPAGB/PPN stars can produce a significant deviation in ISEDs for
young SSPs, but almost no effect for old SSPs in the visible and
infra-red regions. (b) The integrated spectrum of NGC 6838 can be
fitted by a SSP with solar metallicity and an age of $\sim$
12.6\,Gyr. (c) When comparing our synthetic Lick/IDS
absorption-line indices with the values in the literatures and
observational data, we find that our models are generally in
agreement with other models and match real globular clusters in
most index-index diagrams. $TiO_{\rm 1}$ is, however,
systematically smaller.

At last using the merit function $F$, we find the that spectral
indices of the central region of M32 can be fitted with a single
stellar population with an age of $\sim$ 6.5\,Gyr and a solar-like
metallicity with $F=1.00$. Applying the age and the metallicity of
M32 to some index-index diagrams, we see that $H_\beta - Fe5015$
and $H_\beta - Fe5782$ are better diagrams to estimate directly
the age and metallicity from index-index plots.

\section*{ACKNOWLEDGMENTS}
We acknowledge the generous support provided by the Chinese
Natural Science Foundation (Grant No. 10303006, 19925312 \&
10273020), the Chinese Academy of Sciences(KJCX2-SW-T06), Yunnan
Natural Science Foundation (Grant No. 2002A0020Q) and by the 973
scheme (NKBRSF G1999075406). We are deeply indebted to Dr. Lejeune
for making his BaSeL-2.0 model available to us. We also thank Dr
Ph. Podsiadlowski for discussions.

\appendix
\newpage

\onecolumn{
\section[]{\\Table 1: The absorption index for SSPs with the KTG IMF
(stadard models)}}
\begin{longtable}{rrrrrrrrrrrr}
 \hline
 \hline
 $\scriptstyle {\rm Age}$ &
 $\scriptstyle CN_1$ &
 $\scriptstyle CN_2$ &
 $\scriptstyle Ca4277$ &
 $\scriptstyle G4300$ &
 $\scriptstyle Fe4383$ &
 $\scriptstyle Ca4455$ &
 $\scriptstyle Fe4531$ &
 $\scriptstyle Fe4668$ &
 $\scriptstyle H_\beta$ &
 $\scriptstyle Fe5015$ &
 $\scriptstyle Mg_1$ \\
 $\scriptstyle {\rm (Gyr)}$ &
 $\scriptstyle {\rm (mag)}$ &
 $\scriptstyle {\rm (mag)}$ &
 $\scriptstyle {\rm (\AA)}$ &
 $\scriptstyle {\rm (\AA)}$ &
 $\scriptstyle {\rm (\AA)}$ &
 $\scriptstyle {\rm (\AA)}$ &
 $\scriptstyle {\rm (\AA)}$ &
 $\scriptstyle {\rm (\AA)}$ &
 $\scriptstyle {\rm (\AA)}$ &
 $\scriptstyle {\rm (\AA)}$ &
 $\scriptstyle {\rm (mag)}$ \\
&$\scriptstyle Mg_2$ &
 $\scriptstyle Mg_b$ &
 $\scriptstyle Fe5270$ &
 $\scriptstyle Fe5335$ &
 $\scriptstyle Fe5406$ &
 $\scriptstyle Fe5709$ &
 $\scriptstyle Fe5782$ &
 $\scriptstyle NaD$ &
 $\scriptstyle TiO_1$ &
 $\scriptstyle TiO_2$ \\
&$\scriptstyle {\rm (mag)}$ &
 $\scriptstyle {\rm (\AA)}$ &
 $\scriptstyle {\rm (\AA)}$ &
 $\scriptstyle {\rm (\AA)}$ &
 $\scriptstyle {\rm (\AA)}$ &
 $\scriptstyle {\rm (\AA)}$ &
 $\scriptstyle {\rm (\AA)}$ &
 $\scriptstyle {\rm (\AA)}$ &
 $\scriptstyle {\rm (mag)}$ &
 $\scriptstyle {\rm (mag)}$ \\

 \hline
 \endfirsthead

 \caption[]{continued}\\

 \hline
 \hline
 $\scriptstyle {\rm Age}$ &
 $\scriptstyle CN_1$ &
 $\scriptstyle CN_2$ &
 $\scriptstyle Ca4277$ &
 $\scriptstyle G4300$ &
 $\scriptstyle Fe4383$ &
 $\scriptstyle Ca4455$ &
 $\scriptstyle Fe4531$ &
 $\scriptstyle Fe4668$ &
 $\scriptstyle H_\beta$ &
 $\scriptstyle Fe5015$ &
 $\scriptstyle Mg_1$ \\
 $\scriptstyle {\rm (Gyr)}$ &
 $\scriptstyle {\rm (mag)}$ &
 $\scriptstyle {\rm (mag)}$ &
 $\scriptstyle {\rm (\AA)}$ &
 $\scriptstyle {\rm (\AA)}$ &
 $\scriptstyle {\rm (\AA)}$ &
 $\scriptstyle {\rm (\AA)}$ &
 $\scriptstyle {\rm (\AA)}$ &
 $\scriptstyle {\rm (\AA)}$ &
 $\scriptstyle {\rm (\AA)}$ &
 $\scriptstyle {\rm (\AA)}$ &
 $\scriptstyle {\rm (mag)}$ \\
&$\scriptstyle Mg_2$ &
 $\scriptstyle Mg_b$ &
 $\scriptstyle Fe5270$ &
 $\scriptstyle Fe5335$ &
 $\scriptstyle Fe5406$ &
 $\scriptstyle Fe5709$ &
 $\scriptstyle Fe5782$ &
 $\scriptstyle NaD$ &
 $\scriptstyle TiO_1$ &
 $\scriptstyle TiO_2$ \\
&$\scriptstyle {\rm (mag)}$ &
 $\scriptstyle {\rm (\AA)}$ &
 $\scriptstyle {\rm (\AA)}$ &
 $\scriptstyle {\rm (\AA)}$ &
 $\scriptstyle {\rm (\AA)}$ &
 $\scriptstyle {\rm (\AA)}$ &
 $\scriptstyle {\rm (\AA)}$ &
 $\scriptstyle {\rm (\AA)}$ &
 $\scriptstyle {\rm (mag)}$ &
 $\scriptstyle {\rm (mag)}$ \\
 \hline
 \endhead

 \multicolumn{12}{r}{continued on the next page}
 \endfoot
 \hline
 \endlastfoot

 \vspace*{1pt} \\
 \multicolumn{12}{c}{$z = 0.0001$} \\
     1.0  &  -0.114  &   0.089  &  -0.321  &  -3.184  &  -2.013  &  -0.026  &  -0.237  &   1.208  &   5.869  &   0.645  &   0.010  \\&   0.020  &   0.447  &  -0.761  &   0.384  &   0.120  &   0.089  &  -0.211  &   1.163  &   0.014  &   0.001  \\
     2.0  &  -0.307  &  -0.098  &  -0.170  &  -1.992  &  -1.267  &   0.070  &   0.139  &   0.762  &   4.652  &   0.959  &   0.004  \\&   0.029  &   0.593  &  -0.334  &   0.504  &   0.207  &   0.118  &  -0.155  &   1.147  &   0.014  &   0.002  \\
     3.0  &  -0.288  &  -0.086  &  -0.091  &  -1.322  &  -0.873  &   0.110  &   0.293  &   0.544  &   4.082  &   1.055  &   0.002  \\&   0.031  &   0.608  &  -0.173  &   0.535  &   0.225  &   0.130  &  -0.141  &   1.118  &   0.013  &   0.001  \\
     4.0  &  -0.271  &  -0.080  &  -0.023  &  -0.790  &  -0.610  &   0.134  &   0.386  &   0.374  &   3.693  &   1.088  &   0.001  \\&   0.030  &   0.568  &  -0.080  &   0.542  &   0.224  &   0.135  &  -0.139  &   1.044  &   0.012  &   0.000  \\
     5.0  &  -0.271  &  -0.087  &   0.007  &  -0.669  &  -0.503  &   0.137  &   0.402  &   0.294  &   3.595  &   1.070  &   0.000  \\&   0.033  &   0.738  &  -0.078  &   0.534  &   0.214  &   0.129  &  -0.144  &   0.995  &   0.012  &  -0.001  \\
     6.0  &  -0.267  &  -0.093  &   0.037  &  -0.546  &  -0.420  &   0.140  &   0.415  &   0.218  &   3.518  &   1.054  &   0.000  \\&   0.034  &   0.796  &  -0.074  &   0.529  &   0.207  &   0.124  &  -0.149  &   0.959  &   0.011  &  -0.002  \\
     7.0  &  -0.269  &  -0.096  &   0.053  &  -0.503  &  -0.387  &   0.142  &   0.421  &   0.191  &   3.480  &   1.044  &   0.000  \\&   0.035  &   0.823  &  -0.070  &   0.530  &   0.205  &   0.121  &  -0.152  &   0.953  &   0.011  &  -0.003  \\
     8.0  &  -0.270  &  -0.099  &   0.065  &  -0.495  &  -0.357  &   0.144  &   0.419  &   0.180  &   3.463  &   1.031  &   0.001  \\&   0.036  &   0.841  &  -0.067  &   0.533  &   0.205  &   0.120  &  -0.155  &   0.953  &   0.011  &  -0.003  \\
     9.0  &  -0.273  &  -0.102  &   0.074  &  -0.509  &  -0.352  &   0.142  &   0.413  &   0.171  &   3.457  &   1.016  &   0.002  \\&   0.036  &   0.848  &  -0.068  &   0.533  &   0.204  &   0.118  &  -0.158  &   0.955  &   0.011  &  -0.003  \\
    10.0  &  -0.273  &  -0.104  &   0.084  &  -0.526  &  -0.347  &   0.141  &   0.405  &   0.165  &   3.444  &   1.000  &   0.003  \\&   0.037  &   0.857  &  -0.067  &   0.535  &   0.204  &   0.116  &  -0.161  &   0.963  &   0.011  &  -0.004  \\
    11.0  &  -0.274  &  -0.105  &   0.094  &  -0.531  &  -0.321  &   0.143  &   0.402  &   0.167  &   3.416  &   0.990  &   0.004  \\&   0.037  &   0.869  &  -0.058  &   0.542  &   0.208  &   0.116  &  -0.162  &   0.974  &   0.010  &  -0.004  \\
    12.0  &  -0.275  &  -0.107  &   0.105  &  -0.513  &  -0.293  &   0.145  &   0.403  &   0.163  &   3.365  &   0.983  &   0.005  \\&   0.038  &   0.872  &  -0.046  &   0.548  &   0.211  &   0.117  &  -0.162  &   0.985  &   0.010  &  -0.004  \\
    13.0  &  -0.274  &  -0.108  &   0.115  &  -0.523  &  -0.284  &   0.145  &   0.397  &   0.166  &   3.346  &   0.970  &   0.006  \\&   0.038  &   0.869  &  -0.042  &   0.552  &   0.213  &   0.116  &  -0.165  &   1.000  &   0.010  &  -0.005  \\
    14.0  &  -0.275  &  -0.109  &   0.121  &  -0.530  &  -0.295  &   0.144  &   0.388  &   0.170  &   3.323  &   0.956  &   0.007  \\&   0.040  &   0.894  &  -0.041  &   0.555  &   0.213  &   0.115  &  -0.168  &   1.013  &   0.010  &  -0.005  \\
    15.0  &  -0.277  &  -0.111  &   0.131  &  -0.511  &  -0.261  &   0.147  &   0.384  &   0.183  &   3.287  &   0.947  &   0.009  \\&   0.041  &   0.900  &  -0.033  &   0.563  &   0.218  &   0.116  &  -0.167  &   1.024  &   0.010  &  -0.005  \\
    16.0  &  -0.276  &  -0.110  &   0.142  &  -0.464  &  -0.227  &   0.149  &   0.380  &   0.187  &   3.231  &   0.935  &   0.010  \\&   0.041  &   0.903  &  -0.028  &   0.568  &   0.220  &   0.116  &  -0.169  &   1.038  &   0.010  &  -0.005  \\
    17.0  &  -0.273  &  -0.109  &   0.159  &  -0.358  &  -0.160  &   0.154  &   0.386  &   0.178  &   3.130  &   0.930  &   0.011  \\&   0.042  &   0.907  &  -0.012  &   0.576  &   0.223  &   0.116  &  -0.170  &   1.053  &   0.010  &  -0.005  \\
    18.0  &  -0.269  &  -0.106  &   0.179  &  -0.219  &  -0.107  &   0.160  &   0.396  &   0.160  &   3.010  &   0.926  &   0.012  \\&   0.044  &   0.944  &   0.005  &   0.584  &   0.226  &   0.115  &  -0.172  &   1.070  &   0.010  &  -0.005  \\
    19.0  &  -0.151  &   0.005  &   0.199  &  -0.063  &  -0.026  &   0.168  &   0.409  &   0.146  &   2.878  &   0.925  &   0.013  \\&   0.046  &   0.983  &   0.027  &   0.596  &   0.232  &   0.116  &  -0.172  &   1.084  &   0.010  &  -0.005  \\
 \vspace*{1pt} \\
 \multicolumn{12}{c}{$z = 0.0003$} \\
     1.0  &  -0.037  &   0.153  &  -0.151  &  -2.139  &  -1.376  &  -0.034  &   0.180  &   0.035  &   5.395  &   0.823  &   0.008  \\&   0.033  &   0.645  &  -0.252  &   0.324  &   0.122  &   0.132  &  -0.061  &   1.109  &   0.019  &   0.012  \\
     2.0  &  -0.219  &  -0.032  &  -0.014  &  -1.058  &  -0.891  &   0.114  &   0.602  &  -0.058  &   4.494  &   1.292  &   0.006  \\&   0.044  &   0.706  &   0.173  &   0.500  &   0.225  &   0.188  &   0.008  &   1.130  &   0.021  &   0.017  \\
     3.0  &  -0.204  &  -0.020  &   0.069  &  -0.315  &  -0.471  &   0.190  &   0.798  &  -0.112  &   3.948  &   1.478  &   0.005  \\&   0.047  &   0.821  &   0.354  &   0.572  &   0.266  &   0.221  &   0.033  &   1.119  &   0.020  &   0.015  \\
     4.0  &  -0.199  &  -0.018  &   0.134  &   0.269  &  -0.136  &   0.247  &   0.938  &  -0.153  &   3.535  &   1.600  &   0.005  \\&   0.048  &   0.882  &   0.480  &   0.618  &   0.291  &   0.246  &   0.049  &   1.098  &   0.018  &   0.012  \\
     5.0  &  -0.204  &  -0.024  &   0.185  &   0.737  &   0.124  &   0.284  &   1.025  &  -0.207  &   3.220  &   1.656  &   0.004  \\&   0.048  &   0.950  &   0.551  &   0.639  &   0.297  &   0.260  &   0.054  &   1.045  &   0.017  &   0.010  \\
     6.0  &  -0.210  &  -0.032  &   0.232  &   1.171  &   0.378  &   0.316  &   1.098  &  -0.262  &   2.911  &   1.697  &   0.003  \\&   0.049  &   0.993  &   0.607  &   0.657  &   0.300  &   0.267  &   0.057  &   1.004  &   0.016  &   0.008  \\
     7.0  &  -0.214  &  -0.038  &   0.250  &   1.314  &   0.478  &   0.321  &   1.107  &  -0.297  &   2.770  &   1.683  &   0.003  \\&   0.050  &   1.005  &   0.615  &   0.659  &   0.296  &   0.265  &   0.055  &   0.979  &   0.015  &   0.006  \\
     8.0  &  -0.219  &  -0.044  &   0.266  &   1.393  &   0.546  &   0.325  &   1.114  &  -0.319  &   2.679  &   1.673  &   0.003  \\&   0.051  &   1.009  &   0.618  &   0.661  &   0.294  &   0.264  &   0.052  &   0.967  &   0.014  &   0.005  \\
     9.0  &  -0.222  &  -0.047  &   0.279  &   1.419  &   0.582  &   0.327  &   1.116  &  -0.337  &   2.624  &   1.662  &   0.004  \\&   0.052  &   1.002  &   0.617  &   0.661  &   0.292  &   0.262  &   0.049  &   0.966  &   0.014  &   0.004  \\
    10.0  &  -0.224  &  -0.050  &   0.287  &   1.397  &   0.721  &   0.326  &   1.112  &  -0.353  &   2.584  &   1.648  &   0.004  \\&   0.053  &   1.119  &   0.612  &   0.659  &   0.289  &   0.260  &   0.045  &   0.964  &   0.013  &   0.003  \\
    11.0  &  -0.229  &  -0.054  &   0.293  &   1.336  &   0.726  &   0.323  &   1.102  &  -0.362  &   2.622  &   1.629  &   0.005  \\&   0.053  &   1.122  &   0.605  &   0.658  &   0.287  &   0.258  &   0.042  &   0.967  &   0.013  &   0.002  \\
    12.0  &  -0.232  &  -0.057  &   0.295  &   1.243  &   0.689  &   0.316  &   1.085  &  -0.372  &   2.682  &   1.606  &   0.006  \\&   0.052  &   1.109  &   0.594  &   0.653  &   0.284  &   0.256  &   0.037  &   0.972  &   0.013  &   0.002  \\
    13.0  &  -0.237  &  -0.061  &   0.293  &   1.121  &   0.642  &   0.309  &   1.064  &  -0.367  &   2.759  &   1.584  &   0.006  \\&   0.052  &   1.100  &   0.585  &   0.653  &   0.284  &   0.255  &   0.036  &   0.978  &   0.013  &   0.001  \\
    14.0  &  -0.241  &  -0.064  &   0.292  &   1.004  &   0.582  &   0.301  &   1.041  &  -0.367  &   2.822  &   1.561  &   0.008  \\&   0.052  &   1.084  &   0.576  &   0.651  &   0.283  &   0.254  &   0.032  &   0.986  &   0.012  &   0.001  \\
    15.0  &  -0.246  &  -0.069  &   0.285  &   0.874  &   0.537  &   0.294  &   1.019  &  -0.351  &   2.880  &   1.541  &   0.009  \\&   0.052  &   1.070  &   0.570  &   0.653  &   0.285  &   0.254  &   0.032  &   0.994  &   0.012  &   0.000  \\
    16.0  &  -0.250  &  -0.072  &   0.282  &   0.765  &   0.487  &   0.287  &   0.995  &  -0.340  &   2.924  &   1.518  &   0.010  \\&   0.052  &   1.058  &   0.561  &   0.651  &   0.284  &   0.254  &   0.028  &   1.005  &   0.012  &   0.000  \\
    17.0  &  -0.255  &  -0.075  &   0.284  &   0.712  &   0.474  &   0.285  &   0.979  &  -0.325  &   2.935  &   1.501  &   0.011  \\&   0.051  &   1.025  &   0.560  &   0.654  &   0.286  &   0.255  &   0.026  &   1.017  &   0.012  &  -0.001  \\
    18.0  &  -0.257  &  -0.076  &   0.289  &   0.716  &   0.476  &   0.288  &   0.973  &  -0.300  &   2.914  &   1.494  &   0.012  \\&   0.053  &   1.052  &   0.567  &   0.663  &   0.292  &   0.256  &   0.027  &   1.027  &   0.012  &  -0.001  \\
    19.0  &  -0.258  &  -0.074  &   0.303  &   0.799  &   0.510  &   0.294  &   0.976  &  -0.289  &   2.838  &   1.492  &   0.014  \\&   0.055  &   1.085  &   0.579  &   0.672  &   0.296  &   0.258  &   0.026  &   1.042  &   0.012  &  -0.001  \\
 \vspace*{1pt} \\
 \multicolumn{12}{c}{$z = 0.001$} \\
     1.0  &  -0.140  &  -0.005  &   0.029  &  -1.217  &  -0.826  &   0.076  &   0.655  &  -0.275  &   5.046  &   1.357  &   0.016  \\&   0.052  &   0.900  &   0.399  &   0.512  &   0.258  &   0.244  &   0.121  &   1.249  &   0.029  &   0.032  \\
     2.0  &  -0.132  &  -0.010  &   0.174  &  -0.460  &  -0.552  &   0.218  &   0.995  &  -0.154  &   4.447  &   1.945  &   0.014  \\&   0.059  &   1.191  &   0.670  &   0.661  &   0.338  &   0.285  &   0.140  &   1.150  &   0.036  &   0.044  \\
     3.0  &  -0.114  &  -0.004  &   0.318  &   0.797  &   0.145  &   0.371  &   1.339  &  -0.092  &   3.679  &   2.245  &   0.013  \\&   0.067  &   1.325  &   0.909  &   0.794  &   0.413  &   0.332  &   0.179  &   1.148  &   0.033  &   0.040  \\
     4.0  &  -0.098  &   0.006  &   0.395  &   1.500  &   0.615  &   0.453  &   1.513  &  -0.023  &   3.214  &   2.325  &   0.014  \\&   0.072  &   1.386  &   1.033  &   0.885  &   0.470  &   0.359  &   0.208  &   1.193  &   0.032  &   0.037  \\
     5.0  &  -0.087  &   0.012  &   0.456  &   2.081  &   1.019  &   0.525  &   1.650  &   0.074  &   2.841  &   2.457  &   0.016  \\&   0.077  &   1.474  &   1.147  &   0.965  &   0.523  &   0.383  &   0.232  &   1.238  &   0.031  &   0.036  \\
     6.0  &  -0.079  &   0.018  &   0.503  &   2.447  &   1.319  &   0.573  &   1.736  &   0.145  &   2.601  &   2.524  &   0.017  \\&   0.082  &   1.545  &   1.223  &   1.030  &   0.567  &   0.397  &   0.250  &   1.290  &   0.030  &   0.034  \\
     7.0  &  -0.075  &   0.020  &   0.531  &   2.722  &   1.487  &   0.600  &   1.786  &   0.151  &   2.423  &   2.566  &   0.018  \\&   0.084  &   1.563  &   1.265  &   1.057  &   0.590  &   0.409  &   0.257  &   1.306  &   0.028  &   0.032  \\
     8.0  &  -0.078  &   0.016  &   0.533  &   2.857  &   1.563  &   0.607  &   1.797  &   0.113  &   2.326  &   2.556  &   0.018  \\&   0.084  &   1.556  &   1.275  &   1.053  &   0.589  &   0.417  &   0.257  &   1.289  &   0.026  &   0.027  \\
     9.0  &  -0.084  &   0.010  &   0.526  &   2.903  &   1.585  &   0.607  &   1.795  &   0.076  &   2.281  &   2.534  &   0.018  \\&   0.084  &   1.531  &   1.274  &   1.041  &   0.581  &   0.421  &   0.256  &   1.274  &   0.024  &   0.023  \\
    10.0  &  -0.089  &   0.005  &   0.515  &   2.897  &   1.570  &   0.601  &   1.784  &   0.032  &   2.264  &   2.506  &   0.018  \\&   0.085  &   1.510  &   1.266  &   1.024  &   0.568  &   0.423  &   0.253  &   1.259  &   0.022  &   0.020  \\
    11.0  &  -0.097  &  -0.002  &   0.501  &   2.846  &   1.531  &   0.593  &   1.769  &  -0.009  &   2.270  &   2.475  &   0.018  \\&   0.084  &   1.471  &   1.255  &   1.005  &   0.554  &   0.423  &   0.250  &   1.247  &   0.021  &   0.018  \\
    12.0  &  -0.108  &  -0.013  &   0.484  &   2.742  &   1.490  &   0.577  &   1.739  &  -0.063  &   2.295  &   2.426  &   0.018  \\&   0.083  &   1.438  &   1.234  &   0.984  &   0.537  &   0.422  &   0.244  &   1.230  &   0.019  &   0.015  \\
    13.0  &  -0.112  &  -0.017  &   0.473  &   2.636  &   1.595  &   0.569  &   1.725  &  -0.087  &   2.328  &   2.404  &   0.018  \\&   0.082  &   1.486  &   1.223  &   0.952  &   0.507  &   0.421  &   0.242  &   1.233  &   0.019  &   0.014  \\
    14.0  &  -0.118  &  -0.021  &   0.470  &   2.504  &   1.541  &   0.559  &   1.705  &  -0.114  &   2.409  &   2.379  &   0.019  \\&   0.082  &   1.496  &   1.210  &   0.944  &   0.503  &   0.420  &   0.239  &   1.235  &   0.018  &   0.012  \\
    15.0  &  -0.123  &  -0.025  &   0.466  &   2.348  &   1.473  &   0.547  &   1.681  &  -0.140  &   2.501  &   2.348  &   0.020  \\&   0.082  &   1.505  &   1.197  &   0.938  &   0.500  &   0.420  &   0.236  &   1.241  &   0.017  &   0.011  \\
    16.0  &  -0.130  &  -0.031  &   0.451  &   2.128  &   1.372  &   0.527  &   1.641  &  -0.168  &   2.627  &   2.305  &   0.020  \\&   0.083  &   1.502  &   1.174  &   0.927  &   0.494  &   0.417  &   0.234  &   1.244  &   0.017  &   0.010  \\
    17.0  &  -0.137  &  -0.036  &   0.436  &   1.909  &   1.264  &   0.507  &   1.599  &  -0.197  &   2.739  &   2.264  &   0.021  \\&   0.083  &   1.495  &   1.153  &   0.916  &   0.489  &   0.416  &   0.230  &   1.247  &   0.016  &   0.009  \\
    18.0  &  -0.144  &  -0.042  &   0.421  &   1.703  &   1.181  &   0.490  &   1.563  &  -0.205  &   2.827  &   2.230  &   0.022  \\&   0.083  &   1.494  &   1.139  &   0.910  &   0.487  &   0.416  &   0.228  &   1.254  &   0.016  &   0.008  \\
    19.0  &  -0.152  &  -0.048  &   0.414  &   1.561  &   1.123  &   0.480  &   1.535  &  -0.208  &   2.881  &   2.201  &   0.022  \\&   0.083  &   1.499  &   1.130  &   0.908  &   0.487  &   0.416  &   0.226  &   1.262  &   0.015  &   0.007  \\
 \vspace*{1pt} \\
 \multicolumn{12}{c}{$z = 0.004$} \\
     1.0  &  -0.182  &  -0.110  &   0.197  &  -0.812  &  -0.327  &   0.252  &   1.117  &  -0.009  &   5.412  &   2.266  &   0.012  \\&   0.067  &   1.324  &   0.896  &   0.695  &   0.316  &   0.319  &   0.214  &   1.283  &   0.038  &   0.046  \\
     2.0  &  -0.126  &  -0.071  &   0.370  &   0.988  &   0.521  &   0.526  &   1.644  &   0.460  &   4.130  &   3.058  &   0.014  \\&   0.083  &   1.599  &   1.311  &   0.963  &   0.514  &   0.465  &   0.305  &   1.224  &   0.036  &   0.042  \\
     3.0  &  -0.089  &  -0.044  &   0.496  &   2.228  &   1.315  &   0.708  &   1.965  &   0.905  &   3.256  &   3.465  &   0.019  \\&   0.096  &   1.834  &   1.575  &   1.163  &   0.652  &   0.540  &   0.366  &   1.320  &   0.034  &   0.039  \\
     4.0  &  -0.064  &  -0.026  &   0.580  &   3.017  &   1.861  &   0.826  &   2.160  &   1.198  &   2.739  &   3.711  &   0.024  \\&   0.105  &   1.980  &   1.741  &   1.294  &   0.746  &   0.589  &   0.406  &   1.394  &   0.032  &   0.038  \\
     5.0  &  -0.049  &  -0.014  &   0.641  &   3.528  &   2.231  &   0.906  &   2.290  &   1.394  &   2.423  &   3.878  &   0.028  \\&   0.112  &   2.087  &   1.853  &   1.383  &   0.812  &   0.625  &   0.434  &   1.451  &   0.032  &   0.037  \\
     6.0  &  -0.038  &  -0.005  &   0.687  &   3.877  &   2.486  &   0.967  &   2.384  &   1.553  &   2.229  &   4.024  &   0.031  \\&   0.117  &   2.171  &   1.936  &   1.450  &   0.862  &   0.648  &   0.454  &   1.502  &   0.033  &   0.039  \\
     7.0  &  -0.033  &  -0.001  &   0.722  &   4.075  &   2.616  &   1.003  &   2.436  &   1.642  &   2.134  &   4.115  &   0.033  \\&   0.121  &   2.221  &   1.982  &   1.491  &   0.889  &   0.656  &   0.462  &   1.548  &   0.033  &   0.041  \\
     8.0  &  -0.027  &   0.004  &   0.763  &   4.285  &   2.744  &   1.044  &   2.492  &   1.751  &   2.050  &   4.241  &   0.035  \\&   0.124  &   2.282  &   2.034  &   1.536  &   0.917  &   0.662  &   0.468  &   1.601  &   0.035  &   0.044  \\
     9.0  &  -0.023  &   0.006  &   0.790  &   4.444  &   2.857  &   1.071  &   2.532  &   1.799  &   1.978  &   4.294  &   0.037  \\&   0.129  &   2.358  &   2.069  &   1.567  &   0.938  &   0.671  &   0.474  &   1.639  &   0.035  &   0.045  \\
    10.0  &  -0.021  &   0.008  &   0.809  &   4.571  &   2.944  &   1.091  &   2.562  &   1.814  &   1.917  &   4.319  &   0.039  \\&   0.132  &   2.407  &   2.095  &   1.589  &   0.956  &   0.679  &   0.479  &   1.669  &   0.035  &   0.045  \\
    11.0  &  -0.019  &   0.009  &   0.824  &   4.674  &   3.014  &   1.108  &   2.587  &   1.812  &   1.865  &   4.332  &   0.041  \\&   0.135  &   2.441  &   2.116  &   1.607  &   0.970  &   0.687  &   0.482  &   1.696  &   0.035  &   0.044  \\
    12.0  &  -0.018  &   0.009  &   0.836  &   4.762  &   3.068  &   1.121  &   2.606  &   1.791  &   1.818  &   4.333  &   0.043  \\&   0.137  &   2.465  &   2.132  &   1.622  &   0.982  &   0.695  &   0.484  &   1.719  &   0.034  &   0.044  \\
    13.0  &  -0.019  &   0.008  &   0.856  &   4.805  &   3.188  &   1.135  &   2.630  &   1.841  &   1.788  &   4.304  &   0.045  \\&   0.141  &   2.539  &   2.156  &   1.657  &   1.002  &   0.696  &   0.491  &   1.784  &   0.034  &   0.044  \\
    14.0  &  -0.019  &   0.007  &   0.872  &   4.839  &   3.291  &   1.146  &   2.650  &   1.876  &   1.764  &   4.268  &   0.047  \\&   0.144  &   2.598  &   2.175  &   1.686  &   1.018  &   0.697  &   0.497  &   1.840  &   0.034  &   0.043  \\
    15.0  &  -0.020  &   0.007  &   0.883  &   4.868  &   3.371  &   1.153  &   2.666  &   1.882  &   1.742  &   4.222  &   0.049  \\&   0.147  &   2.631  &   2.188  &   1.706  &   1.031  &   0.698  &   0.502  &   1.884  &   0.033  &   0.042  \\
    16.0  &  -0.020  &   0.006  &   0.886  &   4.888  &   3.431  &   1.155  &   2.676  &   1.856  &   1.722  &   4.157  &   0.051  \\&   0.148  &   2.633  &   2.194  &   1.719  &   1.041  &   0.702  &   0.507  &   1.917  &   0.032  &   0.040  \\
    17.0  &  -0.021  &   0.005  &   0.888  &   4.906  &   3.479  &   1.156  &   2.683  &   1.826  &   1.702  &   4.088  &   0.052  \\&   0.151  &   2.640  &   2.197  &   1.729  &   1.049  &   0.704  &   0.511  &   1.947  &   0.030  &   0.037  \\
    18.0  &  -0.022  &   0.004  &   0.887  &   4.889  &   3.526  &   1.152  &   2.686  &   1.792  &   1.702  &   4.011  &   0.054  \\&   0.152  &   2.631  &   2.196  &   1.737  &   1.055  &   0.704  &   0.514  &   1.977  &   0.029  &   0.035  \\
    19.0  &  -0.022  &   0.004  &   0.884  &   4.886  &   3.540  &   1.149  &   2.688  &   1.750  &   1.692  &   3.950  &   0.054  \\&   0.154  &   2.624  &   2.192  &   1.739  &   1.058  &   0.705  &   0.517  &   1.995  &   0.028  &   0.033  \\
 \vspace*{1pt} \\
 \multicolumn{12}{c}{$z = 0.01$} \\
     1.0  &  -0.150  &  -0.088  &   0.305  &   0.242  &   0.491  &   0.504  &   1.575  &   0.664  &   5.007  &   2.855  &   0.014  \\&   0.078  &   1.375  &   1.315  &   1.000  &   0.536  &   0.536  &   0.354  &   1.254  &   0.024  &   0.018  \\
     2.0  &  -0.075  &  -0.032  &   0.583  &   2.560  &   2.011  &   0.888  &   2.253  &   1.925  &   3.286  &   3.857  &   0.027  \\&   0.115  &   2.022  &   1.887  &   1.506  &   0.880  &   0.676  &   0.508  &   1.650  &   0.029  &   0.031  \\
     3.0  &  -0.042  &  -0.005  &   0.708  &   3.605  &   2.833  &   1.063  &   2.537  &   2.467  &   2.606  &   4.212  &   0.039  \\&   0.134  &   2.252  &   2.136  &   1.728  &   1.050  &   0.762  &   0.578  &   1.837  &   0.028  &   0.031  \\
     4.0  &  -0.027  &   0.007  &   0.774  &   4.011  &   3.230  &   1.147  &   2.672  &   2.718  &   2.349  &   4.365  &   0.046  \\&   0.145  &   2.394  &   2.257  &   1.839  &   1.141  &   0.806  &   0.617  &   1.947  &   0.028  &   0.032  \\
     5.0  &  -0.015  &   0.018  &   0.829  &   4.367  &   3.606  &   1.218  &   2.785  &   2.927  &   2.177  &   4.487  &   0.051  \\&   0.154  &   2.501  &   2.354  &   1.929  &   1.214  &   0.839  &   0.646  &   2.038  &   0.028  &   0.033  \\
     6.0  &  -0.007  &   0.025  &   0.871  &   4.590  &   3.831  &   1.269  &   2.865  &   3.068  &   2.061  &   4.573  &   0.056  \\&   0.162  &   2.605  &   2.424  &   1.995  &   1.268  &   0.862  &   0.668  &   2.109  &   0.028  &   0.034  \\
     7.0  &   0.000  &   0.031  &   0.901  &   4.730  &   3.992  &   1.310  &   2.929  &   3.185  &   1.979  &   4.647  &   0.060  \\&   0.168  &   2.674  &   2.480  &   2.048  &   1.314  &   0.879  &   0.687  &   2.169  &   0.029  &   0.036  \\
     8.0  &   0.002  &   0.034  &   0.923  &   4.816  &   4.086  &   1.332  &   2.963  &   3.229  &   1.929  &   4.677  &   0.062  \\&   0.172  &   2.719  &   2.510  &   2.077  &   1.336  &   0.887  &   0.693  &   2.213  &   0.029  &   0.036  \\
     9.0  &   0.007  &   0.038  &   0.950  &   4.963  &   4.225  &   1.364  &   3.011  &   3.305  &   1.857  &   4.728  &   0.065  \\&   0.177  &   2.788  &   2.550  &   2.114  &   1.365  &   0.897  &   0.702  &   2.264  &   0.029  &   0.037  \\
    10.0  &   0.012  &   0.042  &   0.976  &   5.107  &   4.361  &   1.396  &   3.057  &   3.375  &   1.787  &   4.779  &   0.068  \\&   0.182  &   2.852  &   2.590  &   2.151  &   1.393  &   0.906  &   0.710  &   2.312  &   0.030  &   0.038  \\
    11.0  &   0.016  &   0.045  &   1.000  &   5.227  &   4.465  &   1.422  &   3.094  &   3.427  &   1.728  &   4.820  &   0.070  \\&   0.186  &   2.913  &   2.623  &   2.181  &   1.415  &   0.913  &   0.716  &   2.356  &   0.030  &   0.038  \\
    12.0  &   0.018  &   0.047  &   1.016  &   5.305  &   4.549  &   1.442  &   3.121  &   3.455  &   1.684  &   4.848  &   0.072  \\&   0.189  &   2.950  &   2.647  &   2.204  &   1.432  &   0.918  &   0.719  &   2.396  &   0.030  &   0.039  \\
    13.0  &   0.020  &   0.048  &   1.033  &   5.386  &   4.622  &   1.461  &   3.149  &   3.474  &   1.641  &   4.875  &   0.074  \\&   0.192  &   2.978  &   2.672  &   2.226  &   1.449  &   0.924  &   0.722  &   2.432  &   0.030  &   0.039  \\
    14.0  &   0.022  &   0.050  &   1.047  &   5.458  &   4.685  &   1.478  &   3.172  &   3.496  &   1.600  &   4.899  &   0.076  \\&   0.195  &   3.018  &   2.691  &   2.245  &   1.463  &   0.927  &   0.724  &   2.467  &   0.031  &   0.040  \\
    15.0  &   0.022  &   0.050  &   1.058  &   5.506  &   4.731  &   1.492  &   3.190  &   3.498  &   1.569  &   4.918  &   0.078  \\&   0.198  &   3.045  &   2.709  &   2.263  &   1.475  &   0.931  &   0.725  &   2.500  &   0.031  &   0.041  \\
    16.0  &   0.023  &   0.050  &   1.069  &   5.558  &   4.768  &   1.505  &   3.207  &   3.492  &   1.537  &   4.936  &   0.080  \\&   0.200  &   3.065  &   2.725  &   2.277  &   1.486  &   0.935  &   0.726  &   2.529  &   0.031  &   0.041  \\
    17.0  &   0.023  &   0.050  &   1.076  &   5.601  &   4.792  &   1.516  &   3.221  &   3.470  &   1.506  &   4.951  &   0.082  \\&   0.202  &   3.071  &   2.738  &   2.289  &   1.495  &   0.939  &   0.726  &   2.556  &   0.031  &   0.041  \\
    18.0  &   0.018  &   0.045  &   1.100  &   5.607  &   4.892  &   1.527  &   3.242  &   3.521  &   1.496  &   4.922  &   0.084  \\&   0.205  &   3.159  &   2.754  &   2.321  &   1.507  &   0.933  &   0.725  &   2.627  &   0.031  &   0.042  \\
    19.0  &   0.016  &   0.043  &   1.118  &   5.625  &   4.957  &   1.537  &   3.260  &   3.547  &   1.480  &   4.905  &   0.086  \\&   0.207  &   3.219  &   2.766  &   2.344  &   1.517  &   0.929  &   0.725  &   2.680  &   0.031  &   0.043  \\
 \vspace*{2pt} \\
 \multicolumn{12}{c}{$z = 0.02$} \\
     1.0  &  -0.112  &  -0.057  &   0.464  &   1.277  &   1.462  &   0.800  &   2.046  &   2.048  &   4.472  &   3.634  &   0.023  \\&   0.104  &   1.709  &   1.774  &   1.494  &   0.848  &   0.698  &   0.520  &   1.700  &   0.022  &   0.017  \\
     2.0  &  -0.046  &  -0.007  &   0.773  &   3.418  &   3.088  &   1.170  &   2.678  &   3.326  &   2.846  &   4.483  &   0.048  \\&   0.149  &   2.368  &   2.310  &   2.016  &   1.217  &   0.857  &   0.673  &   2.252  &   0.026  &   0.029  \\
     3.0  &  -0.017  &   0.018  &   0.906  &   4.138  &   3.986  &   1.327  &   2.931  &   3.905  &   2.414  &   4.797  &   0.062  \\&   0.175  &   2.698  &   2.540  &   2.236  &   1.392  &   0.933  &   0.746  &   2.513  &   0.027  &   0.034  \\
     4.0  &   0.000  &   0.034  &   0.989  &   4.534  &   4.427  &   1.425  &   3.086  &   4.266  &   2.189  &   4.985  &   0.071  \\&   0.191  &   2.922  &   2.681  &   2.369  &   1.502  &   0.977  &   0.789  &   2.675  &   0.029  &   0.037  \\
     5.0  &   0.014  &   0.047  &   1.057  &   4.836  &   4.774  &   1.504  &   3.208  &   4.542  &   2.023  &   5.128  &   0.079  \\&   0.204  &   3.091  &   2.788  &   2.472  &   1.585  &   1.009  &   0.822  &   2.806  &   0.030  &   0.040  \\
     6.0  &   0.025  &   0.058  &   1.118  &   5.084  &   5.059  &   1.569  &   3.309  &   4.767  &   1.893  &   5.244  &   0.085  \\&   0.214  &   3.230  &   2.875  &   2.554  &   1.653  &   1.032  &   0.847  &   2.916  &   0.031  &   0.042  \\
     7.0  &   0.034  &   0.066  &   1.166  &   5.270  &   5.283  &   1.621  &   3.387  &   4.941  &   1.794  &   5.328  &   0.090  \\&   0.223  &   3.337  &   2.941  &   2.618  &   1.704  &   1.050  &   0.866  &   3.003  &   0.031  &   0.045  \\
     8.0  &   0.042  &   0.074  &   1.206  &   5.423  &   5.473  &   1.666  &   3.455  &   5.090  &   1.712  &   5.403  &   0.094  \\&   0.230  &   3.422  &   2.999  &   2.672  &   1.749  &   1.065  &   0.882  &   3.076  &   0.032  &   0.046  \\
     9.0  &   0.049  &   0.081  &   1.246  &   5.555  &   5.641  &   1.708  &   3.518  &   5.228  &   1.641  &   5.480  &   0.098  \\&   0.237  &   3.492  &   3.052  &   2.722  &   1.787  &   1.077  &   0.897  &   3.148  &   0.033  &   0.049  \\
    10.0  &   0.055  &   0.087  &   1.280  &   5.666  &   5.791  &   1.743  &   3.570  &   5.335  &   1.577  &   5.532  &   0.101  \\&   0.242  &   3.567  &   3.097  &   2.764  &   1.820  &   1.086  &   0.908  &   3.214  &   0.034  &   0.050  \\
    11.0  &   0.061  &   0.093  &   1.311  &   5.754  &   5.917  &   1.775  &   3.616  &   5.428  &   1.526  &   5.581  &   0.104  \\&   0.248  &   3.628  &   3.135  &   2.799  &   1.849  &   1.094  &   0.917  &   3.274  &   0.034  &   0.051  \\
    12.0  &   0.066  &   0.098  &   1.341  &   5.836  &   6.033  &   1.804  &   3.659  &   5.513  &   1.477  &   5.623  &   0.107  \\&   0.252  &   3.684  &   3.172  &   2.834  &   1.877  &   1.101  &   0.925  &   3.332  &   0.034  &   0.051  \\
    13.0  &   0.069  &   0.102  &   1.366  &   5.894  &   6.128  &   1.829  &   3.694  &   5.580  &   1.438  &   5.661  &   0.110  \\&   0.256  &   3.725  &   3.202  &   2.861  &   1.898  &   1.105  &   0.931  &   3.385  &   0.035  &   0.053  \\
    14.0  &   0.073  &   0.105  &   1.387  &   5.944  &   6.210  &   1.851  &   3.725  &   5.632  &   1.403  &   5.688  &   0.112  \\&   0.260  &   3.763  &   3.229  &   2.886  &   1.917  &   1.109  &   0.936  &   3.434  &   0.035  &   0.053  \\
    15.0  &   0.076  &   0.109  &   1.405  &   5.982  &   6.278  &   1.870  &   3.752  &   5.679  &   1.374  &   5.714  &   0.114  \\&   0.263  &   3.797  &   3.252  &   2.907  &   1.934  &   1.112  &   0.940  &   3.480  &   0.036  &   0.054  \\
    16.0  &   0.078  &   0.111  &   1.420  &   6.013  &   6.337  &   1.887  &   3.776  &   5.716  &   1.348  &   5.736  &   0.115  \\&   0.265  &   3.822  &   3.274  &   2.926  &   1.949  &   1.114  &   0.943  &   3.523  &   0.036  &   0.054  \\
    17.0  &   0.080  &   0.114  &   1.429  &   6.029  &   6.385  &   1.900  &   3.795  &   5.744  &   1.327  &   5.752  &   0.117  \\&   0.268  &   3.839  &   3.291  &   2.941  &   1.961  &   1.115  &   0.945  &   3.562  &   0.036  &   0.055  \\
    18.0  &   0.083  &   0.116  &   1.440  &   6.056  &   6.431  &   1.914  &   3.815  &   5.769  &   1.305  &   5.766  &   0.118  \\&   0.270  &   3.854  &   3.309  &   2.957  &   1.972  &   1.117  &   0.947  &   3.600  &   0.037  &   0.056  \\
    19.0  &   0.085  &   0.118  &   1.445  &   6.067  &   6.466  &   1.925  &   3.830  &   5.783  &   1.289  &   5.776  &   0.120  \\&   0.271  &   3.864  &   3.323  &   2.971  &   1.983  &   1.117  &   0.948  &   3.635  &   0.037  &   0.056  \\
 \vspace*{1pt} \\
 \multicolumn{12}{c}{$z = 0.03$} \\
     1.0  &  -0.094  &  -0.044  &   0.565  &   1.841  &   1.973  &   0.975  &   2.299  &   2.910  &   4.163  &   4.076  &   0.031  \\&   0.120  &   1.872  &   2.019  &   1.805  &   1.031  &   0.796  &   0.606  &   2.003  &   0.022  &   0.018  \\
     2.0  &  -0.029  &   0.007  &   0.886  &   3.954  &   3.861  &   1.346  &   2.919  &   4.252  &   2.637  &   4.873  &   0.061  \\&   0.172  &   2.599  &   2.542  &   2.329  &   1.415  &   0.963  &   0.756  &   2.599  &   0.026  &   0.030  \\
     3.0  &  -0.004  &   0.029  &   1.045  &   4.534  &   4.689  &   1.499  &   3.170  &   4.842  &   2.285  &   5.173  &   0.079  \\&   0.202  &   3.026  &   2.771  &   2.564  &   1.597  &   1.036  &   0.832  &   2.917  &   0.029  &   0.039  \\
     4.0  &   0.011  &   0.044  &   1.150  &   4.821  &   5.082  &   1.594  &   3.326  &   5.194  &   2.087  &   5.343  &   0.090  \\&   0.221  &   3.266  &   2.909  &   2.706  &   1.709  &   1.076  &   0.879  &   3.124  &   0.031  &   0.045  \\
     5.0  &   0.024  &   0.057  &   1.242  &   5.078  &   5.427  &   1.677  &   3.459  &   5.497  &   1.923  &   5.484  &   0.100  \\&   0.237  &   3.464  &   3.023  &   2.822  &   1.801  &   1.109  &   0.915  &   3.289  &   0.033  &   0.050  \\
     6.0  &   0.035  &   0.068  &   1.320  &   5.293  &   5.716  &   1.747  &   3.568  &   5.746  &   1.791  &   5.600  &   0.108  \\&   0.250  &   3.624  &   3.114  &   2.914  &   1.874  &   1.132  &   0.942  &   3.423  &   0.034  &   0.054  \\
     7.0  &   0.045  &   0.077  &   1.388  &   5.455  &   5.953  &   1.805  &   3.659  &   5.952  &   1.690  &   5.695  &   0.115  \\&   0.261  &   3.751  &   3.190  &   2.991  &   1.934  &   1.150  &   0.964  &   3.538  &   0.036  &   0.057  \\
     8.0  &   0.053  &   0.085  &   1.442  &   5.574  &   6.142  &   1.852  &   3.732  &   6.119  &   1.611  &   5.772  &   0.120  \\&   0.269  &   3.850  &   3.252  &   3.051  &   1.982  &   1.165  &   0.983  &   3.631  &   0.037  &   0.059  \\
     9.0  &   0.061  &   0.094  &   1.495  &   5.664  &   6.311  &   1.897  &   3.803  &   6.286  &   1.543  &   5.848  &   0.126  \\&   0.278  &   3.937  &   3.311  &   3.110  &   2.028  &   1.177  &   1.002  &   3.721  &   0.038  &   0.063  \\
    10.0  &   0.069  &   0.103  &   1.553  &   5.764  &   6.490  &   1.943  &   3.873  &   6.443  &   1.471  &   5.925  &   0.131  \\&   0.286  &   4.029  &   3.372  &   3.166  &   2.074  &   1.192  &   1.020  &   3.813  &   0.039  &   0.065  \\
    11.0  &   0.076  &   0.110  &   1.602  &   5.840  &   6.638  &   1.981  &   3.932  &   6.569  &   1.412  &   5.984  &   0.135  \\&   0.293  &   4.114  &   3.422  &   3.213  &   2.111  &   1.202  &   1.033  &   3.895  &   0.039  &   0.066  \\
    12.0  &   0.083  &   0.118  &   1.649  &   5.905  &   6.778  &   2.017  &   3.986  &   6.686  &   1.358  &   6.038  &   0.139  \\&   0.300  &   4.184  &   3.468  &   3.256  &   2.145  &   1.211  &   1.046  &   3.972  &   0.040  &   0.067  \\
    13.0  &   0.090  &   0.125  &   1.695  &   5.964  &   6.907  &   2.051  &   4.038  &   6.795  &   1.308  &   6.088  &   0.143  \\&   0.306  &   4.249  &   3.513  &   3.296  &   2.177  &   1.220  &   1.057  &   4.046  &   0.040  &   0.068  \\
    14.0  &   0.096  &   0.132  &   1.734  &   6.002  &   7.022  &   2.081  &   4.084  &   6.889  &   1.267  &   6.128  &   0.147  \\&   0.311  &   4.306  &   3.552  &   3.334  &   2.207  &   1.227  &   1.068  &   4.115  &   0.041  &   0.069  \\
    15.0  &   0.101  &   0.137  &   1.771  &   6.040  &   7.118  &   2.108  &   4.124  &   6.967  &   1.229  &   6.165  &   0.150  \\&   0.316  &   4.352  &   3.586  &   3.363  &   2.230  &   1.231  &   1.075  &   4.179  &   0.041  &   0.070  \\
    16.0  &   0.105  &   0.142  &   1.801  &   6.059  &   7.200  &   2.130  &   4.158  &   7.033  &   1.200  &   6.196  &   0.152  \\&   0.320  &   4.394  &   3.616  &   3.390  &   2.251  &   1.234  &   1.082  &   4.238  &   0.042  &   0.071  \\
    17.0  &   0.110  &   0.148  &   1.832  &   6.083  &   7.283  &   2.153  &   4.193  &   7.101  &   1.168  &   6.222  &   0.155  \\&   0.324  &   4.431  &   3.646  &   3.415  &   2.271  &   1.237  &   1.089  &   4.296  &   0.042  &   0.072  \\
    18.0  &   0.114  &   0.153  &   1.856  &   6.088  &   7.353  &   2.171  &   4.222  &   7.150  &   1.145  &   6.238  &   0.157  \\&   0.327  &   4.456  &   3.672  &   3.437  &   2.288  &   1.240  &   1.095  &   4.351  &   0.042  &   0.072  \\
    19.0  &   0.119  &   0.158  &   1.877  &   6.094  &   7.419  &   2.188  &   4.250  &   7.202  &   1.123  &   6.255  &   0.159  \\&   0.330  &   4.480  &   3.697  &   3.457  &   2.304  &   1.242  &   1.101  &   4.402  &   0.042  &   0.072  \\
\end{longtable}

\bsp
\label{lastpage}

\begin{thebibliography}{99}
\bibitem[\protect\citeauthoryear{Aaronson et al.}{1978}]{aar78}
  Aaronson M., Cohen J. G., Mould J., Malkan M., 1978, ApJ, 223, 824
\bibitem[\protect\citeauthoryear{Allard \& Hauschildt}{1995}]{all95}
  Allard F., Hauschildt P. H., 1995, ApJ, 445, 433
\bibitem[\protect\citeauthoryear{Arimoto}{1996}]{ari96}
  Arimoto N., 1996, APS Conf. Ser. 98, From stars to Galaxies, ed. C. Leitherer, U. Fritze-von Alvensleven, \& J. Huchra, 287
\bibitem[\protect\citeauthoryear{Bertelli et al.}{1994}]{ber94}
  Bertelli G., Bressan A., Chiosi C., Fagotto F., Nasi E., 1994, A\&AS, 106, 275
\bibitem[\protect\citeauthoryear{Bessell et al.}{1989}]{bes89}
  Bessell M. S., Brett J. M., Wood P. R., Scholz M., 1989, A\&AS, 77, 1
\bibitem[\protect\citeauthoryear{Bessell et al.}{1991}]{bes91}
  Bessell M. S., Brett J. M., Scholz M., Wood P. R., 1991, A\&AS, 89, 335
\bibitem[\protect\citeauthoryear{Bressan et al.}{1993}]{bre93}
  Bressan A., Fagotto F., Bertelli G., Chiosi C., 1993, A\&AS, 100, 647
\bibitem[\protect\citeauthoryear{Bressan, Chiosi \& Fagotto}{Bressan et al.}{1994}]{bre94}
  Bressan A., Chiosi C., Fagotto F., 1994, ApJS, 94, 63
\bibitem[\protect\citeauthoryear{Bressan, Chiosi \& Tantalo}{Bressan et al.}{1996}]{bre96}
  Bressan A., Chiosi C., Tantalo R., 1996, A\&A, 311, 425
\bibitem[\protect\citeauthoryear{Bruzual}{1983}]{bru83}
  Bruzual G., 1983, ApJ, 273, 105
\bibitem[\protect\citeauthoryear{Bruzual}{1996}]{bru96}
  Bruzual G., 1996, RevMexA\&A, 4, 61
\bibitem[\protect\citeauthoryear{Bruzual}{2003}]{bru2003}
  Bruzual G., 2003, in Perez-Fournon I., Balcells M., Moreno-Insertis F., Sanchez F., eds, Galaxies at High Redshift. XI Canary
  Islands Winter School of Astrophysics. Cambridge, Cambridge Univ. Press, 185
\bibitem[\protect\citeauthoryear{Bruzual \& Charlot}{1993}]{bru93}
  Bruzual G., Charlot S., 1993, ApJ, 405, 538
\bibitem[\protect\citeauthoryear{Buzzoni}{1989}]{buz89}
  Buzzoni A., 1989, ApJS, 71, 817
\bibitem[\protect\citeauthoryear{Buzzoni}{1995}]{buz95}
  Buzzoni A., 1995, ApJS, 98, 69
\bibitem[\protect\citeauthoryear{Buzzoni, Gariboldi \& Mantegazza}{Buzzoni et al.}{1992}]{buz92}
  Buzzoni A., Gariboldi G., Mantegazza L., 1992, AJ, 103, 1814
\bibitem[\protect\citeauthoryear{Buzzoni, Chincarini \& Molinari}{Buzzoni et al.}{1993}]{buz93}
  Buzzoni A., Chincarini G., Molinari E., 1993, ApJ, 410, 499
\bibitem[\protect\citeauthoryear{Carraro et al.}{1996}]{car96}
  Carraro G., Girardi L., Bressan A., Chiosi C., 1996, A\&A, 305, 849
\bibitem[\protect\citeauthoryear{Chabrier \& Baraffe}{1997}]{cha97}
  Chabrier G., Baraffe I., 1997, A\&AS, 327, 1039
\bibitem[\protect\citeauthoryear{Charbonnel et al.}{1993}]{cha93}
  Charbonnel C., Meynet G., Maeder A., Schaller G., Schaerer D., 1993, A\&A, 101, 415
\bibitem[\protect\citeauthoryear{del Burgo et al.}{2001}]{del2001}
  del Burgo C., Peletier R. F., Vazdekis A., Arribas S., Mediavilla E., MNRAS, 2001, 321, 227
\bibitem[\protect\citeauthoryear{Eggleton}{1971}]{egg71}
  Eggleton P. P., 1971, MNRAS, 151, 351
\bibitem[\protect\citeauthoryear{Eggleton}{1972}]{egg72}
  Eggleton P. P., 1972, MNRAS, 156, 361
\bibitem[\protect\citeauthoryear{Eggleton}{1973}]{egg73}
  Eggleton P. P., 1973, MNRAS, 163, 279
\bibitem[\protect\citeauthoryear{Fagotto et al.}{1994a}]{fag94a}
  Fagotto F., Bressan A., Bertelli G., Chiosi C., 1994a, A\&AS, 104, 365
\bibitem[\protect\citeauthoryear{Fagotto et al.}{1994b}]{fag94b}
  Fagotto F., Bressan A., Bertelli G., Chiosi C., 1994b, A\&AS, 105, 29
\bibitem[\protect\citeauthoryear{Fagotto et al.}{1994c}]{fag94c}
  Fagotto F., Bressan A., Bertelli G., Chiosi C., 1994c, A\&AS, 105, 39
\bibitem[\protect\citeauthoryear{Fulks et al.}{1994}]{ful94}
  Fluks M. A., Plez B., Th$\acute e$ P. S., De Winter D., Westerlund B. E., Steenman H. C., 1994, A\&AS, 105, 311
\bibitem[\protect\citeauthoryear{Frogel, Persson \& Cohen}{Frogel et al.}{1980}]{fro80}
  Frogel J. A., Persson S. E., Cohen J. G., 1980, ApJ, 240, 785
\bibitem[\protect\citeauthoryear{Gunn \& Stryker}{1983}]{gun83}
  Gunn J. E., Stryker L. L., 1983, ApJS, 52, 121
\bibitem[\protect\citeauthoryear{Gorgas et al.}{1993}]{gor93}
  Gorgas J., Faber S. M., Burstein D., Gonz\'alez J. J., Courteau S., Prosser C., 1993, ApJS, 86, 153
\bibitem[\protect\citeauthoryear{Green \& Demarque}{1987}]{gre87}
  Green E. W., Demarque P., King C. R., 1987, The Revised Yale Isochrones and Luminosity functions (New Haven: Yale University Observatory)
\bibitem[\protect\citeauthoryear{Grillmair et al.}{1996}]{gri96}
  Grillmair C. J., Lauer T. R., Worthey G., Faber S. M., Freedman W. L., Madore B.
  F., Ajhar E. A., Baum W. A., et al., 1996, AJ, 112, 1975
\bibitem[\protect\citeauthoryear{Han, Podsiadlowski \& Eggleton}{Han et al.}{1994}]{han94}
  Han Z., Podsiadlowski P., Eggleton P. P., 1994, MNRAS, 270, 121
\bibitem[\protect\citeauthoryear{Hurley, Pols \& Tout}{Hurley et al.}{2000}]{hur2000}
  Hurley J. R., Pols O. R., Tout C. A., 2000, MNRAS, 315, 543
\bibitem[\protect\citeauthoryear{Iben \& Renzini}{1983}]{ibe83}
  Iben I. Jr., Renzini A., 1983, ARA\&A, 21, 271
\bibitem[\protect\citeauthoryear{Jones \& Worthey}{1995}]{jon95}
  Jones L. A., Worthey G., 1995, ApJ, 446, L31
\bibitem[\protect\citeauthoryear{Kodama \& Arimoto}{1997}]{kod97}
  Kodama T., Arimoto N., 1997, A\&A, 320, 41
\bibitem[\protect\citeauthoryear{Kroupa, Tout \& Gilmore}{Kroupa et al.}{1993}]{kro93}
  Kroupa P., Tout C. A., Gilmore G., 1993, MNRAS, 262, 545 ({\bf KTG})
\bibitem[\protect\citeauthoryear{Kurth, Alvensleben \& Fricke}{Kurth et al.}{1999}]{kur99}
  Kurth O. M., Alvensleben U. F.-v., Fricke K. J., 1999, A\&AS, 138, 19 ({\bf KAF})
\bibitem[\protect\citeauthoryear{Kurucz}{1992}]{kur92}
  Kurucz R. L., 1992, The Stellar Populations of Galaxies, IAU Symp. 149, ed. B. Barbuy \& A. Renzini (Dordrech: Kluwer), 225
\bibitem[\protect\citeauthoryear{Lejeune}{1997}]{lej97a}
  Lejeune T., 1997a, Ph.D. thesis, Univ. Louis Pasteur, Strasbourg, France \& Astron. Institute of Basel University, Basel, Switzerland
\bibitem[\protect\citeauthoryear{Lejeune, Cuisinier \& Buser}{Lejeune et al.}{1997}]{lej97b}
  Lejeune T., Cuisinier F., Buser R., 1997b, A\&AS, 125, 229
\bibitem[\protect\citeauthoryear{Lejeune, Cuisinier \& Buser}{Lejeune et al.}{1998}]{lej98}
  Lejeune T., Cuisinier F., Buser R., 1998, A\&AS, 130, 65
\bibitem[\protect\citeauthoryear{Mould}{1978}]{mou78}
  Mould J. R., 1978, ApJ, 220, 434
\bibitem[\protect\citeauthoryear{Mowlavi et al.}{1998}]{mow98}
  Mowlavi N., Schaerer D., Meynet G., Bernasconi P. A., Charbonnel C., Maeder A., 1998, A\&AS, 128, 471
\bibitem[\protect\citeauthoryear{O'Connell}{1976}]{oco76}
  O'Connell R. W., 1976, ApJ, 206, 370
\bibitem[\protect\citeauthoryear{O'Connell}{1986}]{oco86}
  O'Connell R. W., 1986, in Stellar Populations, ed. C. A. Norman, A. Renzini, \& M. Tosi (Cambridge: Cambridge Univ.
  Press), 167
\bibitem[\protect\citeauthoryear{O'Connell}{1994}]{oco94}
  O'Connell R. W., 1994, in Genzel T., Harris A. I, eds, NATO Adv.
  Sci. Inst. Ser. C 445, The Neclei of Normal Galaxies: Lessons from the Galactic Center, Dordrecht:
  Kluwer, P. 255
\bibitem[\protect\citeauthoryear{Peletier}{1989}]{pel89}
  Peletier R. F., 1989, Ph.D. thesis, Univ. of Groningen, The Netherland
\bibitem[\protect\citeauthoryear{Pols et al.}{1995}]{pol95}
  Pols O. R., Tout C. A., Eggleton P. P., Han Z., 1995, MNRAS, 274, 964
\bibitem[\protect\citeauthoryear{Pols et al.}{1998}]{pol98}
  Pols O. R., Schr$\ddot o$der K. P., Hurley J. R., Tout C. A., Eggleton P. P., 1998, MNRAS, 298, 525
\bibitem[\protect\citeauthoryear{Reimers}{1975}]{rei75}
  Reimers D., 1975, Mem. Soc. R. Sci. Li$\grave e$ge, 6e Ser., 8, 369
\bibitem[\protect\citeauthoryear{Renzini}{1981}]{ren81}
  Renzini A., 1981, in Effects of Mass Loss on Stellar Evolution, ed. C. Chiosi \& R. Stalio (Dordrecht: Reidel), 319
\bibitem[\protect\citeauthoryear{Renzini}{1986}]{ren86}
  Renzini A., 1986, in Stellar Populations, ed. C. A. Norman, A. Renzini, \& M. Tosi (Cambridge: Cambridge Univ. Press), 213
\bibitem[\protect\citeauthoryear{Rose}{1994}]{ros94}
  Rose J. A., 1994, AJ, 106, 107
\bibitem[\protect\citeauthoryear{Salpeter}{1955}]{sal55}
  Salpeter E. E., 1955, ApJ, 121, 161
\bibitem[\protect\citeauthoryear{Schaller et al.}{1992}]{sch92}
  Schaller G., Schaerer D., Meynet G., Maeder A., 1992, A\&AS, 96, 269
\bibitem[\protect\citeauthoryear{Tantalo et al.}{1996}]{tan96}
  Tantalo R., Chiosi C., Bressan A., Fagotto F., 1996, A\&A, 311, 361
\bibitem[\protect\citeauthoryear{Tinsley}{1972a}]{tin72a}
  Tinsley B. M., 1972a, A\&A, 20, 383
\bibitem[\protect\citeauthoryear{Tinsley}{1972b}]{tin72b}
  Tinsley B. M., 1972b, ApJ, 178, 39L
\bibitem[\protect\citeauthoryear{Tinsley \& Gunn}{1976}]{tin76}
  Tinsley B. M., Gunn J. E., 1976, ApJ, 203, 52
\bibitem[\protect\citeauthoryear{Trager et al.}{1998}]{tra98}
  Trager S. C., Worthey G., Faber S. M., Burstein D., Gonzalez J. J., 1998, ApJS, 116, 1
\bibitem[\protect\citeauthoryear{Tripicco \& Bell}{1992}]{tri92}
  Tripicco M. J., Bell R. A., 1992, AJ, 103, 1285
\bibitem[\protect\citeauthoryear{VandenBerg}{1985}]{van85a}
  VandenBerg D. A., 1985, ApJS, 58, 711
\bibitem[\protect\citeauthoryear{VandenBerg}{1992}]{van92}
  VandenBerg D. A., 1992, ApJ, 391, 685
\bibitem[\protect\citeauthoryear{VandenBerg \& Bell}{1985}]{van85b}
  VandenBerg D. A., Bell R. A., 1985, ApJS, 58, 561
\bibitem[\protect\citeauthoryear{VandenBerg \& Laskarides}{1987}]{van87}
  VandenBerg D. A., Laskarides P. G., 1987, ApJS, 64, 103
\bibitem[\protect\citeauthoryear{Vazdekis}{1999}]{vaz99a}
  Vazdekis A., 1999, ApJ, 513, 224
\bibitem[\protect\citeauthoryear{Vazdekis}{2001}]{vaz2001}
  Vazdekis A., 2001, Ap\&SS, 276, 839
\bibitem[\protect\citeauthoryear{Vazdekis \& Arimoto}{1999}]{vaz99b}
  Vazdekis A., Arimoto N., 1999, ApJ, 525, 144
\bibitem[\protect\citeauthoryear{Vazdekis et al.}{1996}]{vaz96}
  Vazdekis A., Casuso E., Peletier R. F., Beckman J. E., 1996, ApJS,
  106, 307 ({\bf VCPB})
\bibitem[\protect\citeauthoryear{Worthey}{1992}]{wor92}
  Worthey G., 1992, Ph.D. thesis, Univ. of California, Santa Cruz
\bibitem[\protect\citeauthoryear{Worthey}{1994}]{wor94a}
  Worthey G., 1994, ApJS, 95, 107 ({\bf GW})
\bibitem[\protect\citeauthoryear{}{astro.wsu.edu/worthey/dial/dial\_a\_model.html}]{wor2002}
  Worthey G., astro.wsu.edu/worthey/dial/dial\_a\_model.html
\bibitem[\protect\citeauthoryear{Worthey \& Ottaviani}{1997}]{wor97}
  Worthey G., Ottaviani D. L., 1997, ApJS, 111, 377
\bibitem[\protect\citeauthoryear{Worthey et al.}{1994}]{wor94b}
  Worthey G., Faber S. M., Gonzalez J. J., Burstein D., 1994, ApJS, 94, 687
\bibitem[\protect\citeauthoryear{Zhang et al.}{2002}]{zha2002}
  Zhang F., Han Z., Li L., Hurley J. R., 2002, MNRAS, 334, 833 (Paper I)

\end{thebibliography}
\end{document}